\documentstyle[prd,preprint,tighten,aps,eqsecnum,%
amssymb,newlfont,epsfig]{revtex}
\begin{document}
\draft
\preprint{
\begin{tabular}{r}
UWThPh-1997-11\\
DFTT 26/97\\
hep-ph/9705300
\end{tabular}
}
\title{CONSTRAINTS ON LONG-BASELINE NEUTRINO OSCILLATION PROBABILITIES
AND CP ASYMMETRIES FROM NEUTRINO OSCILLATION DATA}
\author{S.M. Bilenky}
\address{Joint Institute for Nuclear Research, Dubna, Russia, and\\
Technion, Physics Department, 32000 Haifa, Israel.}
\author{C. Giunti}
\address{INFN, Sezione di Torino, and Dipartimento di Fisica Teorica,
Universit\`a di Torino,\\
Via P. Giuria 1, I--10125 Torino, Italy.}
\author{W. Grimus}
\address{Institute for Theoretical Physics, University of Vienna,\\
Boltzmanngasse 5, A--1090 Vienna, Austria.}
\maketitle
\begin{abstract}
We consider long-baseline neutrino oscillations 
in the framework of two schemes with mixing of four 
massive neutrinos which can accommodate
all the existing indications in
favour of neutrino mixing.
Within these schemes,
we derive bounds on
the oscillation probabilities and
the CP-odd neutrino--antineutrino
asymmetries in long-baseline experiments.
Using the results of
short-baseline neutrino oscillation experiments, we obtain rather
strong upper bounds on the long-baseline probabilities 
$1-P^{(\mathrm{LBL})}_{\bar\nu_e \to \bar\nu_e}$
and
$P^{(\mathrm{LBL})}_{\stackrel{\makebox[0pt][l]
{$\hskip-4pt\scriptscriptstyle(-)$}}{\nu_\mu}
\to\stackrel{\makebox[0pt][l]
{$\hskip-4pt\scriptscriptstyle(-)$}}{\nu_e}}$.
Nevertheless,
the projected sensitivities of the MINOS and ICARUS
experiments are better than our bounds.
We also show that 
there are no corresponding constraints for
$
\stackrel{\makebox[0pt][l]
{$\hskip-3pt\scriptscriptstyle(-)$}}{\nu_{\mu}}
\to\stackrel{\makebox[0pt][l]
{$\hskip-3pt\scriptscriptstyle(-)$}}{\nu_{\mu}}
$
and
$
\stackrel{\makebox[0pt][l]
{$\hskip-3pt\scriptscriptstyle(-)$}}{\nu_{\mu}}
\to\stackrel{\makebox[0pt][l]
{$\hskip-3pt\scriptscriptstyle(-)$}}{\nu_{\tau}}
$
long-baseline oscillations
and that the CP-odd
asymmetry in the latter channel
can reach the maximal value
$ 2/3\sqrt{3} $
allowed
by the unitarity of the mixing matrix.
Some schemes with mixing of
three neutrinos are also considered.
\end{abstract}


\narrowtext

\section{Introduction}
\label{Introduction}

The search for neutrino oscillations remains a central issue
of neutrino experiments.
In several short-baseline (SBL) experiments, 
which are sensitive to relatively large values of the
neutrino mass-squared difference
$\delta{m}^2$
($ \delta{m}^2 \gtrsim 0.1 \, \mathrm{eV}^2 $),
different
oscillation channels are investigated
at present:
$
\stackrel{\makebox[0pt][l]
{$\hskip-3pt\scriptscriptstyle(-)$}}{\nu_{\mu}}
\to\stackrel{\makebox[0pt][l]
{$\hskip-3pt\scriptscriptstyle(-)$}}{\nu_{e}}
$
(LSND \cite{LSND},
KARMEN \cite{KARMEN},
CCFR \cite{CCFR96},
NOMAD \cite{NOMADmuel});
$
\stackrel{\makebox[0pt][l]
{$\hskip-3pt\scriptscriptstyle(-)$}}{\nu_{\mu}}
\to\stackrel{\makebox[0pt][l]
{$\hskip-3pt\scriptscriptstyle(-)$}}{\nu_{\tau}}
$
(CHORUS \cite{CHORUS},
NOMAD \cite{NOMAD} and CCFR \cite{CCFR95}). 
The long-baseline (LBL) neutrino oscillation experiments
which will operate in the near future
will be sensitive to much smaller values of
$\delta{m}^2$
($ \delta{m}^2 \gtrsim 10^{-3} \, \mathrm{eV}^2 $).
Long-baseline experiments with reactor antineutrinos
are starting to take data (CHOOZ \cite{CHOOZ})
or are under preparation
(Palo Verde \cite{PaloVerde}).
Long-baseline experiments with accelerator neutrinos
(KEK--Super-Kamiokande \cite{KEKSK},
Fermilab--Soudan \cite{MINOS},
CERN--Gran Sasso \cite{ICARUS})
are under preparation.

No indications in favour of 
neutrino oscillations have been found
in many experiments with terrestrial neutrinos which have been
done in the past
(see the reviews in Ref.\cite{Boehm-Vannucci}).
On the other hand,
such indications have been found in
all solar neutrino experiments
(Homestake,
Kamiokande,
GALLEX
and SAGE
\cite{solarexp}).
The suppression of
the detected event rates
with respect to those
predicted by the
Standard Solar Model (SSM)
\cite{SSM}
can be explained
with neutrino mixing.
In the case of resonant MSW transitions
\cite{MSW},
it was found
\cite{SOLMSW}
that the oscillation parameters
$\delta{m}^2$
and
$\sin^{2}2\vartheta$
($\vartheta$ is the mixing angle) have the following values:
\begin{equation}
3 \times 10^{-6}
\lesssim
\delta{m}^{2}
\lesssim
1.2 \times 10^{-5} \, \mathrm{eV}^2
\;,
\qquad
4 \times 10^{-3}
\lesssim
\sin^{2}2\vartheta
\lesssim
1.2 \times 10^{-2}
\;.
\label{11}
\end{equation}
The solar neutrino puzzle can also be solved by invoking vacuum
neutrino oscillations \cite{barger},
in which case $\delta m^2 \sim 10^{-10} \mathrm{eV}^2$.

Indications in favour of
$
\stackrel{\makebox[0pt][l]
{$\hskip-3pt\scriptscriptstyle(-)$}}{\nu_{\mu}}
\to\stackrel{\makebox[0pt][l]
{$\hskip-3pt\scriptscriptstyle(-)$}}{\nu_{x}}
$
oscillations ($x\neq\mu$)
have been found also in the
Kamiokande
\cite{Kamiokande-atmospheric},
IMB \cite{IMB}
and
Soudan
\cite{Soudan}
atmospheric neutrino experiments.
From the analysis of the Kamiokande data
the following allowed ranges
for the oscillation parameters were obtained \cite{Kamiokande-atmospheric}:
\begin{eqnarray}
5 \times 10^{-3}
\lesssim
\delta m^2
\lesssim
3 \times 10^{-2} \, \mbox{eV}^2
\;,
\null & \null
\qquad
0.7
\lesssim
\sin^2 2\vartheta
\lesssim
1
\null & \null
\qquad
(\nu_{\mu} \leftrightarrows \nu_{\tau})
\;,
\label{32}
\\
7 \times 10^{-3}
\lesssim
\delta m^2
\lesssim
8 \times 10^{-2} \, \mbox{eV}^2
\;,
\null & \null
\qquad
0.6
\lesssim
\sin^2 2\vartheta
\lesssim
1
\null & \null
\qquad
(\nu_{\mu} \leftrightarrows \nu_{e})
\;.
\label{33}
\end{eqnarray}

Finally,
indications in favour of
$ \bar\nu_\mu \to \bar\nu_e $
oscillations
have been found recently in the LSND experiment
\cite{LSND},
in which antineutrinos originating from the decays of 
$\mu^+$'s 
at rest were detected.
From the analysis of the data of this experiment
and the negative results of other SBL experiments
(in particular,
the Bugey \cite{Bugey95} and BNL E776 \cite{BNLE776}
experiments),
it follows that
\begin{equation}
0.3
\lesssim
\delta{m}^2
\lesssim
2.2 \, \mathrm{eV}^2
\;,
\qquad
10^{-3}
\lesssim
\sin^2 2\vartheta
\lesssim
4 \times 10^{-2}
\;.
\label{051}
\end{equation}

All the above-mentioned hints for neutrino oscillations can be
accommodated
in schemes with four neutrinos.
These schemes require a sterile neutrino
field $(\alpha = s)$ in addition to the usual flavour fields with flavours
$\alpha = e, \mu, \tau$.
The specific nature of such 4-neutrino schemes
has recently been derived from the data
\cite{BGG96,OY96}.

The purpose of the present paper is to consider LBL neutrino
oscillations in the light of the 4-neutrino schemes favoured by the data.
In this context we will discuss
the bounds on the LBL oscillation
probabilities and on
the CP-odd neutrino--antineutrino asymmetries
that can be obtained from the results of the
SBL oscillation experiments.
We will also make an excursion to some
3-neutrino schemes.

Let us stress that CP violation is one of the important problems which
can be tackled in LBL experiments.
If CP is violated in the
lepton sector,
then the neutrino mixing matrix is complex and the
probabilities of
$\nu_\alpha \to \nu_\beta$
and
$\bar\nu_\alpha \to \bar\nu_\beta$
transitions
$(\alpha\neq\beta)$
are different.
The observation of CP violation in neutrino oscillations
will be very important for the understanding
of the nature of neutrino
mixing and the nature of CP violation.
In the framework of mixing of
three massive neutrinos,
possible effects of CP violation in
LBL experiments were considered in Refs.\cite{TA96,AR96,MN96}.
In the
present paper we will derive bounds on
the CP-odd asymmetries in
LBL experiments in the framework of
the schemes with mixing of
four massive neutrinos which take into account all
the existing neutrino
oscillation data.
We will present general methods for obtaining such bounds
from the results of SBL experiments. 
These methods will be applied to some
mixing schemes of three neutrinos as well.
In the context of the 4-neutrino schemes,
we will show that 
sizable CP-odd asymmetries can be expected only in 
$
\stackrel{\makebox[0pt][l]
{$\hskip-3pt\scriptscriptstyle(-)$}}{\nu_{\mu}}
\to\stackrel{\makebox[0pt][l]
{$\hskip-3pt\scriptscriptstyle(-)$}}{\nu_{\tau}}
$
transitions.

The plan of the paper is as follows.
In Section
\ref{Neutrino oscillations and CP violation}
we will review the
basics of neutrino oscillations and CP violation.
The 4-neutrino schemes
favoured by the data and the relevant LBL formulas are
introduced in Section
\ref{Four massive neutrinos}.
Section
\ref{Constraints on the long-baseline probabilities}
is devoted to the discussion of
the bounds on the LBL oscillation probabilities
which follow from the
4-neutrino schemes and the SBL data.
The same is done for the CP
violation parameters in section
\ref{CP violation in the schemes with four neutrinos}.
Finally, we make a digression to
3-neutrino schemes in Section
\ref{Three massive neutrinos}
and we formulate our conclusions in Section
\ref{Conclusions}.
The three Appendices contain some derivations and discussions 
of the bounds used in the main body of the paper.

\section{Neutrino oscillations and CP violation}
\label{Neutrino oscillations and CP violation}

In accordance with the neutrino mixing hypothesis
(see, for example, Refs.\cite{BP78,BP87,CWKim}),
a left-handed neutrino field
$\nu_{{\alpha}L}$
is a mixture of the left-handed
components
$\nu_{kL}$
of the (Dirac or Majorana) fields of neutrinos
with definite masses
$m_k$:
\begin{equation}
\nu_{{\alpha}L}
=
\sum_{k}
U_{{\alpha}k}
\,
\nu_{kL}
\quad \mbox{with} \quad
\alpha=e,\mu,\tau,s,\ldots
\label{05}
\end{equation}
where $U$ is the unitary mixing matrix and $\nu_{sL}$ a sterile
neutrino field.
The mixing in Eq.(\ref{05})
implies that
the probabilities of transitions
of neutrinos (antineutrinos) with momentum $p$ at a distance $L$ of
the neutrino detector from the neutrino source
are given by
\begin{eqnarray}
&&
P_{\nu_\alpha\to\nu_\beta}
=
\left|
\sum_{k}
U_{{\beta}k}
\,
U_{{\alpha}k}^{*}
\,
\exp\!\left(
- i \, \frac{ \Delta{m}^{2}_{k1} \, L }{ 2 \, p }
\right)
\right|^2
\;,
\label{052}
\\
&&
P_{\bar\nu_\alpha\to\bar\nu_\beta}
=
\left|
\sum_{k}
U_{{\beta}k}^{*}
\,
U_{{\alpha}k}
\,
\exp\!\left(
- i \, \frac{ \Delta{m}^{2}_{k1} \, L }{ 2 \, p }
\right)
\right|^2
\;,
\label{053}
\end{eqnarray}
where 
$ \Delta{m}^{2}_{k1} \equiv m_k^2 - m_1^2 $
(we take
$ m_1 < m_2 < \ldots $).
From Eqs.(\ref{052}) and (\ref{053})
it follows that the transition probabilities of
neutrinos and antineutrinos are connected by the relation
\begin{equation}
P_{\nu_\alpha\to\nu_\beta}
=
P_{\bar\nu_\beta\to\bar\nu_\alpha}
\;.
\label{054}
\end{equation}
This relation reflects CPT invariance.

If CP invariance in the lepton sector holds,
then there are phase
conventions such that
in the case of Dirac neutrinos
we have
\begin{equation}
U_{{\alpha}k}
=
U_{{\alpha}k}^{*}
\;,
\label{0551}
\end{equation}
whereas
in the case of Majorana neutrinos
we have
\begin{equation}
U_{{\alpha}k}
= -
U_{{\alpha}k}^{*}
\,
\eta_{k}
\label{0552}
\end{equation}
where
$\eta_{k}={\pm}i$
is the CP parity\footnote{
The CP parities of Majorana neutrinos could 
be important for neutrinoless double beta-decay;
for example, if
the $\nu_k$'s
have different CP parities,
their contributions
to the amplitude
of neutrinoless double-beta decay
could cancel each other
\cite{DBB}.}
of the Majorana neutrino with mass $m_k$
(see, for example, Ref.\cite{BP87}).
It is obvious that the CP parities
$\eta_{k}$
do not enter in the expressions
for the transitions amplitudes.
Hence,
in both the Dirac and Majorana cases,
CP invariance
implies that
\cite{BHP80-Doi81}
\begin{equation}
P_{\nu_\alpha\to\nu_\beta}
=
P_{\bar\nu_\alpha\to\bar\nu_\beta}
\;.
\label{056}
\end{equation}

Let us introduce the CP-odd asymmetries
\begin{equation}
D_{\alpha;\beta}
\equiv
P_{\nu_\alpha\to\nu_\beta}
-
P_{\bar\nu_\alpha\to\bar\nu_\beta}
\;.
\label{057}
\end{equation}
From CPT invariance it follows that
\begin{equation}
D_{\alpha;\beta}
=
-
D_{\beta;\alpha}
\;.
\label{058}
\end{equation}
Furthermore,
from the unitarity of the mixing matrix
we have
\begin{equation}
\sum_{\beta\neq\alpha}
D_{\alpha;\beta}
=
0
\;.
\label{059}
\end{equation}
We observe that in the case of transitions
among three flavour states
($\nu_e$, $\nu_\mu$, $\nu_\tau$)
the CP asymmetries satisfy the relations
\cite{CPrelations} 
\begin{equation}\label{CPrelations}
D_{e;\mu}
=
D_{\tau;e}
=
D_{\mu;\tau}
\;,
\label{060}
\end{equation}
which follow from Eqs.(\ref{058}) and (\ref{059}).

In the general case of mixing
of an arbitrary number of massive neutrinos,
the asymmetries are given by
\begin{equation}
D_{\alpha;\beta}
=
4
\sum_{k>j}
\mbox{Im}\!\left[
U_{{\alpha}j}
\,
U_{{\beta}j}^{*}
\,
U_{{\alpha}k}^{*}
\,
U_{{\beta}k}
\right]
\sin
\frac{ \Delta{m}^{2}_{kj} \, L }{ 2 \, p }
\;.
\label{061}
\end{equation}
Therefore,
CP violation in the lepton sector can be observed
only if at least one of the oscillating terms in the transition
probabilities does not vanish
due to the averaging over the neutrino energy spectrum
and the size of the neutrino source and detector.

\section{Four massive neutrinos}
\label{Four massive neutrinos}

All existing indications in favour of neutrino
oscillations can be accommodated
by a scheme with mixing of
four massive neutrinos
\cite{four,BGKP,BGG96,OY96}.
In Ref.\cite{BGG96}
we have shown that from the six possible spectral schemes
of four massive neutrinos,
which correspond to three
different scales of mass-squared differences
$\Delta{m}^{2}_{kj}$,
only two schemes
are compatible with the results of all experiments.
In these two schemes
the four neutrino masses
are divided in two pairs of close masses
separated by a gap of
$ \sim 1 \, \mathrm{eV} $:
\begin{equation}
\mbox{(A)}
\qquad
\underbrace{
\overbrace{m_1 < m_2}^{\mathrm{atm}}
\ll
\overbrace{m_3 < m_4}^{\mathrm{solar}}
}_{\mathrm{LSND}}
\qquad \mbox{and} \qquad
\mbox{(B)}
\qquad
\underbrace{
\overbrace{m_1 < m_2}^{\mathrm{solar}}
\ll
\overbrace{m_3 < m_4}^{\mathrm{atm}}
}_{\mathrm{LSND}}
\;.
\label{AB}
\end{equation}
In scheme A,
$\Delta{m}^{2}_{21}$
is relevant
for the explanation of the atmospheric neutrino anomaly
and
$\Delta{m}^{2}_{43}$
is relevant
for the suppression of solar $\nu_e$'s.
In scheme B,
the roles of
$\Delta{m}^{2}_{21}$
and
$\Delta{m}^{2}_{43}$
are reversed.

We will consider first
SBL neutrino oscillations
in the framework of these two schemes.
Let us note above all that it is impossible to reveal 
the effects of CP violation in these experiments
which are sensitive to
$ \Delta{m}^{2}_{41} \gtrsim 0.1 \, \mathrm{eV}^2 $.
Furthermore, both schemes A and B give the same oscillation
probabilities. Therefore, it is impossible to distinguish A from B in
SBL oscillation experiments either.
In fact,
since in these experiments
\begin{equation} \label{ass}
\frac{ \Delta{m}^{2}_{21} \, L }{ 2 \, p } \ll 1
\qquad
\mbox{and}
\qquad
\frac{ \Delta{m}^{2}_{43} \, L }{ 2 \, p } \ll 1
\;,
\end{equation}
the probabilities of
$ \nu_\alpha \to \nu_\beta $
and
$ \bar\nu_\alpha \to \bar\nu_\beta $
transitions are equal and are given by
\cite{BGKP}
\begin{equation} \label{P12}
P^{(\mathrm{SBL})}_{\stackrel{\makebox[0pt][l]
{$\hskip-3pt\scriptscriptstyle(-)$}}{\nu_{\alpha}}
\to\stackrel{\makebox[0pt][l]
{$\hskip-3pt\scriptscriptstyle(-)$}}{\nu_{\beta}}}
=
\frac{1}{2}
\,
A_{\alpha;\beta}
\left( 1 - \cos\frac{ \Delta{m}^{2} \, L }{ 2 \, p } \right)
\;,
\end{equation}
with
$ \Delta{m}^{2} \equiv \Delta{m}^{2}_{41} = m^2_4 - m^2_1 $
and the oscillation amplitude
\begin{equation} \label{A12}
A_{\alpha;\beta}
=
4 \left| \sum_{k=1,2} U_{{\beta}k} U_{{\alpha}k}^{*} \right|^2
=
4 \left| \sum_{k=3,4} U_{{\beta}k} U_{{\alpha}k}^{*} \right|^2
\;.
\end{equation}
The survival probabilities of neutrinos and antineutrinos
are given by
\begin{equation} \label{P11}
P^{(\mathrm{SBL})}_{\stackrel{\makebox[0pt][l]
{$\hskip-3pt\scriptscriptstyle(-)$}}{\nu_{\alpha}}
\to\stackrel{\makebox[0pt][l]
{$\hskip-3pt\scriptscriptstyle(-)$}}{\nu_{\alpha}}}
=
1 - \sum_{\beta\neq\alpha}
P^{(\mathrm{SBL})}_{\stackrel{\makebox[0pt][l]
{$\hskip-3pt\scriptscriptstyle(-)$}}{\nu_{\alpha}}
\to\stackrel{\makebox[0pt][l]
{$\hskip-3pt\scriptscriptstyle(-)$}}{\nu_{\beta}}}
=
1
-
\frac{1}{2}
\,
B_{\alpha;\alpha}
\left( 1 - \cos\frac{ \Delta{m}^{2} \, L }{ 2 \, p } \right)
\;,
\end{equation}
where
\begin{equation}
\begin{array}{rcl} \displaystyle
B_{\alpha;\alpha}
=
\sum_{\beta\neq\alpha}
A_{\alpha;\beta}
& = & \displaystyle
4
\left( \sum_{k=1,2} |U_{{\alpha}k}|^2 \right)
\left( 1 - \sum_{k=1,2} |U_{{\alpha}k}|^2 \right)
\\ \displaystyle
& = & \displaystyle
4
\left( \sum_{k=3,4} |U_{{\alpha}k}|^2 \right)
\left( 1 - \sum_{k=3,4} |U_{{\alpha}k}|^2 \right)
\;.
\end{array}
\label{B11}
\end{equation}

Long-baseline neutrino oscillation experiments are
planned to be sensitive
to the ``atmospheric neutrino range''
$
10^{-3} \, \mathrm{eV}^2
\lesssim
\Delta{m}^{2}_{kj}
\lesssim
10^{-1} \, \mathrm{eV}^2
$.
In scheme A,
the probabilities of
$ \nu_\alpha \to \nu_\beta $
and
$ \bar\nu_\alpha \to \bar\nu_\beta $
transitions
in LBL experiments
are given by
\begin{eqnarray}
&&
P^{(\mathrm{LBL,A})}_{\nu_\alpha\to\nu_\beta}
=
\left|
U_{\beta1}
\,
U_{\alpha1}^{*}
+
U_{\beta2}
\,
U_{\alpha2}^{*}
\,
\exp\!\left(
- i
\frac{ \Delta{m}^{2}_{21} \, L }{ 2 \, p }
\right)
\right|^2
+
\left|
\sum_{k=3,4}
U_{{\beta}k}
\,
U_{{\alpha}k}^{*}
\right|^2
\;,
\label{plba1}
\\
&&
P^{(\mathrm{LBL,A})}_{\bar\nu_\alpha\to\bar\nu_\beta}
=
\left|
U_{\beta1}^{*}
\,
U_{\alpha1}
+
U_{\beta2}^{*}
\,
U_{\alpha2}
\,
\exp\!\left(
- i
\frac{ \Delta{m}^{2}_{21} \, L }{ 2 \, p }
\right)
\right|^2
+
\left|
\sum_{k=3,4}
U_{{\beta}k}^{*}
\,
U_{{\alpha}k}
\right|^2
\;.
\label{plba2}
\end{eqnarray}
These formulas have been obtained from Eqs.(\ref{052}) and (\ref{053}),
respectively,
taking into account the fact that in LBL experiments
$ \Delta{m}^{2}_{43} L / 2 p \ll 1 $
and dropping the terms proportional to 
the cosines of phases much larger
than $2\pi$
($ \Delta{m}^{2}_{kj} L / 2 p \gg 2\pi $
for $k=3,4$ and $j=1,2$).
Such terms do not contribute to the oscillation
probabilities averaged over the
neutrino energy spectrum.
The transition probabilities
in scheme B
ensue from the expressions
(\ref{plba1}) and (\ref{plba2})
with the change
\begin{equation}
1 \, , \, 2
\leftrightarrows
3 \, , \, 4
\;.
\label{0721}
\end{equation}

From Eqs.(\ref{plba1}), (\ref{plba2}) and (\ref{0721})
it follows that the LBL CP-odd asymmetries
$D^{(\mathrm{LBL})}_{\alpha;\beta}$
in the schemes A and B are given by
\begin{eqnarray}
&&
D^{(\mathrm{LBL,A})}_{\alpha;\beta}
=
I^{(\mathrm{A})}_{\alpha\beta}
\,
\sin
\frac{ \Delta{m}^{2}_{21} \, L }{ 2 \, p }
\;,
\label{073}
\\
&&
D^{(\mathrm{LBL,B})}_{\alpha;\beta}
=
I^{(\mathrm{B})}_{\alpha\beta}
\,
\sin
\frac{ \Delta{m}^{2}_{43} \, L }{ 2 \, p }
\;,
\label{074}
\end{eqnarray}
with
the oscillation amplitudes
\begin{equation}
I^{(\mathrm{A})}_{\alpha\beta}
\equiv
4
\,
\mbox{Im}\!\left[
U_{{\alpha}1}
\,
U_{{\beta}1}^{*}
\,
U_{{\alpha}2}^{*}
\,
U_{{\beta}2}
\right]
\;,
\qquad
I^{(\mathrm{B})}_{\alpha\beta}
\equiv
4
\,
\mbox{Im}\!\left[
U_{{\alpha}3}
\,
U_{{\beta}3}^{*}
\,
U_{{\alpha}4}^{*}
\,
U_{{\beta}4}
\right]
\;.
\label{iab}
\end{equation}

We will now discuss
the constraints on the neutrino oscillation parameters
which follow from the existing results
of SBL experiments.
In the disappearance SBL
reactor and accelerator experiments
no indication in favour of neutrino oscillations
was found
(see the reviews in Ref.\cite{Boehm-Vannucci}).
At any fixed value of $\Delta{m}^{2}$
in the range
\begin{equation}
0.1 \, \mathrm{eV}^2
\leq \Delta{m}^{2} \leq
10^3 \, \mathrm{eV}^2
\;,
\label{075}
\end{equation}
from the
exclusion plots of the
Bugey \cite{Bugey95}
$\bar\nu_e$
disappearance experiment
and of the
CDHS \cite{CDHS84} and CCFR \cite{CCFR84}
$\stackrel{(-)}{\nu}_\mu$
disappearance experiments we have
\begin{equation} \label{B0}
B_{\alpha;\alpha} \leq B_{\alpha;\alpha}^{0}
\qquad
( \alpha = e , \mu )
\;.
\end{equation}

Let us define the quantities
$c_{\alpha}$
(with $\alpha=e,\mu,\tau$)
in the two schemes A and B as
\begin{eqnarray}
(\mbox{A})
\null & \null \qquad \null & \null
c_{\alpha}
\equiv
\sum_{k=1,2} |U_{{\alpha}k}|^2
\;,
\label{dca}
\\
(\mbox{B})
\null & \null \qquad \null & \null
c_{\alpha}
\equiv
\sum_{k=3,4} |U_{{\alpha}k}|^2
\;.
\label{dcb}
\end{eqnarray}
Taking also into account the results
of the solar neutrino experiments and
the results of the atmospheric neutrino experiments,
in both schemes A and B,
the quantities
$c_{e}$ and $c_{\mu}$
are constrained by
\cite{BGG96}
\begin{equation}
c_{e} \leq a^{0}_{e}
\;,
\qquad
c_{\mu} \geq 1 - a^{0}_{\mu}
\;,
\label{cc}
\end{equation}
where
\begin{equation} \label{a0}
a^{0}_{\alpha} = \frac{1}{2}
\left(1-\sqrt{1-B_{\alpha;\alpha}^{0}}\,\right)
\quad (\alpha = e,\mu)
\;.
\end{equation}
The values of
$a^{0}_{e}$
and
$a^{0}_{\mu}$
which have
obtained from the exclusion plots of
the Bugey \cite{Bugey95},
CDHS \cite{CDHS84} and CCFR \cite{CCFR84}
experiments are
given in Fig.1 of Ref.\cite{BBGK}.
For values of
$\Delta{m}^{2}$
in the range (\ref{075})
$a^{0}_{e}$
is small
($ a^{0}_e \lesssim 4 \times 10^{-2} $),
and
$a^{0}_{\mu}$
is small for
$
\Delta{m}^{2} \gtrsim 0.3 \, \mathrm{eV}^2
$
($ a^{0}_\mu \lesssim 10^{-1} $).
In the following
we will use also the limits on the amplitudes
$A_{\alpha;\beta}$
which can be obtained from exclusion plots of the
BNL E734
\cite{BNLE734},
BNL E776
\cite{BNLE776}
and
CCFR
\cite{CCFR96}
$
\stackrel{\makebox[0pt][l]
{$\hskip-3pt\scriptscriptstyle(-)$}}{\nu_{\mu}}
\to\stackrel{\makebox[0pt][l]
{$\hskip-3pt\scriptscriptstyle(-)$}}{\nu_{e}}
$
appearance experiments
and of the
FNAL E531
\cite{FNALE531}
and
CCFR
\cite{CCFR95}
$
\stackrel{\makebox[0pt][l]
{$\hskip-3pt\scriptscriptstyle(-)$}}{\nu_{\mu}}
\to\stackrel{\makebox[0pt][l]
{$\hskip-3pt\scriptscriptstyle(-)$}}{\nu_{\tau}}
$
appearance experiments:
\begin{equation} \label{A0}
A_{\alpha;\beta} \leq A_{\alpha;\beta}^{0}
\;,
\end{equation}
with
$\alpha=\mu$
and
$\beta=e,\tau$.
The results of the LSND
\cite{LSND}
$ \bar\nu_\mu \to \bar\nu_e $
appearance experiment
will be also
taken into account
(in particular, the allowed range (\ref{051}) of
$\Delta m^2$).

Since scheme B emerges from scheme A by the substitution (\ref{0721}) and
since we will derive bounds on the LBL
oscillation probabilities
$
P^{(\mathrm{LBL,A})}_{\stackrel{\makebox[0pt][l]
{$\hskip-3pt\scriptscriptstyle(-)$}}{\nu_{\alpha}}
\to\stackrel{\makebox[0pt][l]
{$\hskip-3pt\scriptscriptstyle(-)$}}{\nu_{\beta}}}
$,
$
P^{(\mathrm{LBL,B})}_{\stackrel{\makebox[0pt][l]
{$\hskip-3pt\scriptscriptstyle(-)$}}{\nu_{\alpha}}
\to\stackrel{\makebox[0pt][l]
{$\hskip-3pt\scriptscriptstyle(-)$}}{\nu_{\beta}}}
$
and on the CP-odd parameters
$I^{(\mathrm{A})}_{\alpha\beta}$,
$I^{(\mathrm{B})}_{\alpha\beta}$
as functions of $A_{\alpha;\beta}$, $c_\alpha$ and $c_\beta$,
it is
evident that such bounds apply equally to both schemes A and B by
virtue of the definitions (\ref{A12}), (\ref{dca}) and
(\ref{dcb}). Consequently, when dealing with such bounds 
we will omit the superscripts A, B
indicating the specific scheme.

\section{Constraints on the long-baseline probabilities}
\label{Constraints on the long-baseline probabilities}

We will consider first the limits on the
LBL oscillation probabilities
which can be obtained from
the results of the
SBL oscillation experiments.
The Cauchy--Schwarz inequality
implies for scheme A that
\begin{equation}
\left|
\sum_{k=1,2}
U_{{\beta}k}
\,
U_{{\alpha}k}^{*}
\,
\exp\!\left(
- i
\frac{ \Delta{m}^{2}_{k1} \, L }{ 2 \, p }
\right)
\right|^2
\leq
c_{\alpha}
\,
c_{\beta}
\;.
\label{111}
\end{equation}
Using this inequality and the definition
(\ref{dca})
of $c_\alpha$,
we find from the LBL probabilities in
Eqs.(\ref{plba1}) and (\ref{plba2})
that the survival probabilities
$P^{(\mathrm{LBL})}_{\nu_{\alpha}\to\nu_{\alpha}}$
and
$P^{(\mathrm{LBL})}_{\bar\nu_{\alpha}\to\bar\nu_{\alpha}}$
are bounded by
\begin{equation}
\left( 1 - c_{\alpha} \right)^2
\leq
P^{(\mathrm{LBL})}_{\stackrel{\makebox[0pt][l]
{$\hskip-3pt\scriptscriptstyle(-)$}}{\nu_{\alpha}}
\to\stackrel{\makebox[0pt][l]
{$\hskip-3pt\scriptscriptstyle(-)$}}{\nu_{\alpha}}}
\leq
c_{\alpha}^2
+
\left( 1 - c_{\alpha} \right)^2
\;.
\label{paa}
\end{equation}
As explained at the end of the last section,
these bounds are
scheme-independent. 
In order to obtain bounds on the 
LBL transition probabilities
$P^{(\mathrm{LBL})}_{\nu_{\alpha}\to\nu_{\beta}}$
and
$P^{(\mathrm{LBL})}_{\bar\nu_{\alpha}\to\bar\nu_{\beta}}$,
we take into account the definition
(\ref{A12}) of $A_{\alpha;\beta}$
and the inequality (\ref{111}).
When inserted into Eqs.(\ref{plba1}) and (\ref{plba2})
they imply
\begin{equation}
\frac{1}{4}
\,
A_{\alpha;\beta}
\leq
P^{(\mathrm{LBL})}_{\stackrel{\makebox[0pt][l]
{$\hskip-3pt\scriptscriptstyle(-)$}}{\nu_{\alpha}}
\to\stackrel{\makebox[0pt][l]
{$\hskip-3pt\scriptscriptstyle(-)$}}{\nu_{\beta}}}
\leq
c_{\alpha}
\,
c_{\beta}
+
\frac{1}{4}
\,
A_{\alpha;\beta}
\;.
\label{pab1}
\end{equation}
The bounds (\ref{paa}) and (\ref{pab1})
are the basis of the following
considerations for the different oscillation channels in LBL
experiments.

The smallness of $c_e$
(see Eq.(\ref{cc}))
implies that the electron neutrino has a
small mixing with the neutrinos whose mass-squared difference is
responsible for the oscillations of atmospheric and LBL neutrinos
($\nu_1$, $\nu_2$ in scheme A and $\nu_3$, $\nu_4$ in scheme
B). Hence, the probability of transitions of atmospheric and
LBL electron neutrinos into other states is suppressed.
Indeed, taking into account the constraint $ c_e \leq a^{0}_{e} $, 
the lower bound on
$P^{(\mathrm{LBL})}_{\bar\nu_{e}\to\bar\nu_{e}}$ and the upper bounds on
$P^{(\mathrm{LBL})}_{\bar\nu_{\mu}\to\bar\nu_{e}}$ which we will derive
are rather strict.

Let us discuss first the bounds on
the LBL survival probability
$P^{(\mathrm{LBL})}_{\bar\nu_{e}\to\bar\nu_{e}}$.
We will compare these bounds with the expected
sensitivity of the CHOOZ \cite{CHOOZ} and Palo Verde \cite{PaloVerde}
LBL reactor experiments.
Taking into account the constraint (\ref{cc}) on
$c_{e}$,
Eq.(\ref{paa})
implies that
in both schemes A and B
\begin{equation}
1
-
P^{(\mathrm{LBL})}_{\stackrel{\makebox[0pt][l]
{$\hskip-3pt\scriptscriptstyle(-)$}}{\nu_{e}}
\to\stackrel{\makebox[0pt][l]
{$\hskip-3pt\scriptscriptstyle(-)$}}{\nu_{e}}}
\leq
a^{0}_{e}
\left( 2 - a^{0}_{e} \right)
\;.
\label{081}
\end{equation}
The curve corresponding
to this limit
obtained from the 90\% CL exclusion plot of the Bugey
\cite{Bugey95}
experiment is shown
in Fig.\ref{fig1}
(solid line).
The expected sensitivities
of the LBL reactor neutrino experiments
CHOOZ and Palo Verde
are also shown in
Fig.\ref{fig1}
as the dash-dotted and dash-dot-dotted vertical lines, respectively.
These expected sensitivities with respect to 
$1-P^{(\mathrm{LBL})}_{\bar \nu_e \to \bar \nu_e}$
have been extracted by us
from the figures presented in Refs.\cite{CHOOZ,PaloVerde}
showing the sensitivity of the respective experiments
in the
$\sin^{2}2\vartheta$--$\delta{m}^{2}$
plane,
using the fact that for high values of
$\delta{m}^{2}$
each experiment is sensitive only to
the averaged survival probability
$
P^{(\mathrm{LBL})}_{\bar\nu_{e}\to\bar\nu_{e}}
=
1
-
\frac{1}{2}
\sin^{2}2\vartheta
$.
Thus,
the vertical lines in Fig.\ref{fig1} correspond to 
$ \frac{1}{2} \sin^2 2\vartheta $ at high $\delta m^2$.

Figure \ref{fig1}
shows that,
in the framework of the two schemes (\ref{AB}) with four neutrinos,
which allow to accommodate all 
the indications in favour of neutrino oscillations,
the existing data 
put rather strong limitations on the probability of
LBL transitions of
$\nu_e$ into other states
(for
$ \Delta{m}^2 \gtrsim 3 \, \mathrm{eV}^2 $
the upper bound for
$ 1 - P^{(\mathrm{LBL})}_{\bar\nu_{e}\to\bar\nu_{e}} $
is close to the border of the region of sensitivity of the
CHOOZ experiment,
whereas for
$ \Delta{m}^2 \lesssim 3 \, \mathrm{eV}^2 $
it is much smaller).

The shadowed region in Fig.\ref{fig1}
corresponds to the range (\ref{051}) of $\Delta{m}^2$
allowed at 90\% CL by the results of the LSND
and all the other SBL experiments.
It can be seen that
the LSND signal indicates an upper bound for
$1-P^{(\mathrm{LBL})}_{\bar \nu_e \to \bar \nu_e}$
of about
$ 5 \times 10^{-2} $,
smaller than the expected sensitivities of
the CHOOZ and Palo Verde experiments.

Let us stress that,
in the framework of the schemes
under consideration,
the smallness of $c_e$
is a consequence of the solar neutrino problem.
Consider for example scheme A.
The probability of solar neutrinos
to survive is given by
\begin{equation}
P^{(\mathrm{sun},\mathrm{A})}_{\nu_e\to\nu_e}
=
\sum_{k=1,2} |U_{ek}|^4
+
\left( 1 - \sum_{k=1,2} |U_{ek}|^2 \right)^2
P^{(3;4)}_{\nu_e\to\nu_e}
\;,
\label{Psol}
\end{equation}
where $P^{(3;4)}_{\nu_e\to\nu_e}$ is the
survival probability
due to the mixing of $\nu_e$ with $\nu_3$ and $\nu_4$,
depending on the
small mass-squared difference
$\Delta{m}^{2}_{43}$.
From the results of SBL reactor experiments 
it follows that the quantity
$
c_e
\equiv
\displaystyle
\sum_{k=1,2} |U_{ek}|^2
$
can be small or large (close to one).
In order to
have the energy dependence of
the survival probability
$P^{(\mathrm{sun},\mathrm{A})}_{\nu_e\to\nu_e}$
and the suppression of the flux of solar $\nu_e$'s
that are required for the
explanation of the data of solar neutrino experiments,
we must choose a small value of 
$c_e$.
In this case,
the
survival probability of $\bar\nu_e$'s
in LBL reactor experiments
is close to one.

Let us now discuss the bounds on
$
\stackrel{\makebox[0pt][l]
{$\hskip-3pt\scriptscriptstyle(-)$}}{\nu_{\mu}}
\to\stackrel{\makebox[0pt][l]
{$\hskip-3pt\scriptscriptstyle(-)$}}{\nu_{e}}
$
transitions in LBL accelerator experiments.
We will compare these bounds with
the expected sensitivities of the
KEK--Super-Kamiokande (KEK--SK) \cite{KEKSK},
Fermilab--Soudan (MINOS) \cite{MINOS}
and
CERN--Gran Sasso (ICARUS) \cite{ICARUS}
experiments.

Taking into account the constraints (\ref{cc}) on
$c_{e}$ and (\ref{A0}) on
$A_{\mu;e}$,
Eq.(\ref{pab1})
implies that
in both schemes A and B
\begin{equation}
P^{(\mathrm{LBL})}_{\stackrel{\makebox[0pt][l]
{$\hskip-3pt\scriptscriptstyle(-)$}}{\nu_{\mu}}
\to\stackrel{\makebox[0pt][l]
{$\hskip-3pt\scriptscriptstyle(-)$}}{\nu_{e}}}
\leq
a^{0}_{e}
+
\frac{1}{4}
\,
A_{\mu;e}^{0}
\;.
\label{042}
\end{equation}
The curve corresponding
to this limit
obtained from the 90\% CL exclusion plots of the Bugey
\cite{Bugey95}
experiment for
$a^{0}_{e}$
and
of the
BNL E734
\cite{BNLE734},
BNL E776
\cite{BNLE776}
and
CCFR
\cite{CCFR95}
experiments
for
$A_{\mu;e}^{0}$
is shown
in Fig.\ref{fig2}
(long-dashed line).
For a comparison,
we have shown the expected sensitivities
of the LBL accelerator neutrino experiments
KEK--SK \cite{KEKSK},
MINOS \cite{MINOS}
and
ICARUS \cite{ICARUS}
(the dotted, dash-dotted and dash-dot-dotted
vertical lines, respectively).
These sensitivities have been obtained
from the figures presented in Refs.\cite{KEKSK,MINOS,ICARUS}
showing the sensitivities of the respective experiments
in the
$\sin^{2}2\vartheta$--$\delta{m}^{2}$
plane with the method explained in the context of LBL
reactor experiments.

The conservation of probability
and Eq.(\ref{paa}) lead to a further upper bound:
\begin{equation}
P^{(\mathrm{LBL})}_{\nu_\alpha \rightarrow \nu_\beta}
\leq
1
-
P^{(\mathrm{LBL})}_{\nu_\alpha \rightarrow \nu_\alpha}
\leq
c_{\alpha} \left( 2 - c_{\alpha} \right)
\qquad (\alpha \neq \beta)\,.
\label{pab2}
\end{equation}
CPT invariance (see Eq.(\ref{054}))
and the fact that the bound (\ref{paa}) is valid
for antineutrinos as well give the same upper bound as
Eq.(\ref{pab2}) for the opposite transition
$\nu_\beta \rightarrow \nu_\alpha$:
\begin{equation}
P^{(\mathrm{LBL})}_{\nu_\beta \rightarrow \nu_\alpha} =
P^{(\mathrm{LBL})}_{\bar\nu_\alpha \rightarrow \bar\nu_\beta}
\leq
1
-
P^{(\mathrm{LBL})}_{\bar\nu_\alpha \rightarrow \bar\nu_\alpha}
\leq
c_{\alpha} \left( 2 - c_{\alpha} \right)
\qquad (\alpha \neq \beta)\,.
\label{pab3}
\end{equation}
Finally, these two equations hold evidently also for antineutrinos.
The solid curve in Fig.\ref{fig2}
represents the limit
\begin{equation}
P^{(\mathrm{LBL})}_{\stackrel{\makebox[0pt][l]
{$\hskip-3pt\scriptscriptstyle(-)$}}{\nu_{\mu}}
\to\stackrel{\makebox[0pt][l]
{$\hskip-3pt\scriptscriptstyle(-)$}}{\nu_{e}}}
\leq
a^{0}_{e}
\left( 2 - a^{0}_{e} \right)
\label{0421}
\end{equation}
obtained from Eq.(\ref{pab3})
and the constraint (\ref{cc}) on $c_e$.
This bound is better than the bound (\ref{042}) for the
SBL parameter $\Delta m^2 \lesssim 0.4 \,
\mathrm{eV}^2$.

The darkly shadowed area
in Fig.\ref{fig2}
represents the region allowed by the results
of the LSND \cite{LSND} experiment,
taking into account the results of all the other
SBL experiments.
The lower bound on
$
P^{(\mathrm{LBL})}_{\stackrel{\makebox[0pt][l]
{$\hskip-3pt\scriptscriptstyle(-)$}}{\nu_{\mu}}
\to\stackrel{\makebox[0pt][l]
{$\hskip-3pt\scriptscriptstyle(-)$}}{\nu_{e}}}
$,
i.e. the left edge of the shadowed region,
is determined by the results of the LSND experiment,
whereas the upper bound
is given
by the limits obtained before, i.e.
by the most stringent of
the inequalities (\ref{0421}) and (\ref{042})
represented, respectively,
by the solid and long-dashed curves.
Thus,
for each fixed value of $\Delta{m}^2$
we have
\begin{equation}
\frac{1}{4}
\,
A_{\mu;e}^{\mathrm{min}}
\leq
P^{(\mathrm{LBL})}_{\stackrel{\makebox[0pt][l]
{$\hskip-3pt\scriptscriptstyle(-)$}}{\nu_{\mu}}
\to\stackrel{\makebox[0pt][l]
{$\hskip-3pt\scriptscriptstyle(-)$}}{\nu_{e}}}
\leq
\min\!\left(
a^{0}_{e}
\left( 2 - a^{0}_{e} \right)
\, , \,
a^{0}_{e}
+
\frac{1}{4}
\,
A^{0}_{\mu;e}
\right)
\;,
\label{pab12}
\end{equation}
where
$A_{\mu;e}^{\mathrm{min}}$
is the minimal value of
$A_{\mu;e}$
allowed at
90\% CL by the LSND experiment.

Figure \ref{fig2}
shows that,
in the framework of the schemes under consideration,
the sensitivity of the
KEK--SK experiment
may be not sufficient to reveal
LBL
$ \nu_\mu \to \nu_e $ 
oscillations,
whereas the sensitivities of the
MINOS and ICARUS
experiments are considerably better
than the
upper bound for
$P^{(\mathrm{LBL})}_{\nu_{\mu}\to\nu_{e}}$.
It is interesting to observe
that there is also a lower bound
on this transition probability
that follows from the
LSND results (see Eq.(\ref{pab12})).
However,
this lower bound is valid only in the case of
LBL neutrino oscillations in vacuum.
The corrections due to the matter effects
in LBL experiments
make it disappear (see Section\ref{Conclusions}).

The solid curve in Fig.\ref{fig2} is at the same time an upper
bound on 
$
P^{(\mathrm{LBL})}_{\stackrel{\makebox[0pt][l]
{$\hskip-3pt\scriptscriptstyle(-)$}}{\nu_{e}}
\to\stackrel{\makebox[0pt][l]
{$\hskip-3pt\scriptscriptstyle(-)$}}{\nu_{\tau}}}
$.  
This is evident
from Eq.(\ref{pab2}).
On the other hand,
the probability of
$
\stackrel{\makebox[0pt][l]
{$\hskip-3pt\scriptscriptstyle(-)$}}{\nu_{\mu}}
\to\stackrel{\makebox[0pt][l]
{$\hskip-3pt\scriptscriptstyle(-)$}}{\nu_{\tau}}
$
transitions
is not constrained by the results
of SBL experiments.

Finally, a further upper bound on
$
P^{(\mathrm{LBL})}_{\stackrel{\makebox[0pt][l]
{$\hskip-3pt\scriptscriptstyle(-)$}}{\nu_{\alpha}}
\to\stackrel{\makebox[0pt][l]
{$\hskip-3pt\scriptscriptstyle(-)$}}{\nu_{\beta}}}
$ 
for $\alpha \neq \beta$ is gained from
Eq.(\ref{pab1}).
Since
$A_{\alpha;\beta} \leq 4(1-c_\alpha)(1-c_\beta)$,
we have
\begin{equation}\label{bbb}
P^{(\mathrm{LBL})}_{\stackrel{\makebox[0pt][l]
{$\hskip-3pt\scriptscriptstyle(-)$}}{\nu_{\alpha}}
\to\stackrel{\makebox[0pt][l]
{$\hskip-3pt\scriptscriptstyle(-)$}}{\nu_{\beta}}}
\leq
c_\alpha c_\beta + (1-c_\alpha)(1-c_\beta)
\qquad
(\alpha\neq\beta)
\;.
\end{equation}
Obviously, if $c_\alpha = c_\beta = 0$ or 1 is in the allowed range of
these quantities,
then this upper bound is 1 and thus is trivial.
This
leaves only
$\alpha = \mu$ and $\beta = e$
as a non-trivial case,
with
\begin{equation}\label{bbb1}
P^{(\mathrm{LBL})}_{\stackrel{\makebox[0pt][l]
{$\hskip-3pt\scriptscriptstyle(-)$}}{\nu_{\mu}}
\to\stackrel{\makebox[0pt][l]
{$\hskip-3pt\scriptscriptstyle(-)$}}{\nu_{e}}}
\leq
a^0_e + a^0_\mu - 2 a^0_e a^0_\mu \, .
\end{equation}
The short-dashed curve in Fig.\ref{fig2}
shows this limit
with $a^{0}_{e}$  and $a^{0}_{\mu}$
obtained from
the 90\% CL exclusion plots
of the Bugey \cite{Bugey95}
$\bar\nu_{e}\to\bar\nu_{e}$
experiment
and of
the CDHS \cite{CDHS84} and CCFR \cite{CCFR84}
$\bar\nu_{\mu}\to\bar\nu_{\mu}$
experiments,
respectively.
For
$ a^0_\mu \ll a^0_e \ll 1 $
this bound is about half of that given by
Eq.(\ref{0421}).
However,
since $a^0_\mu$ is only small
in the same range of $\Delta{m}^2$ where
$A_{\mu;e}^{0}$ is small,
numerically the bound
(\ref{bbb1})
turns out to be worse than
the bound (\ref{042})
(the long-dashed curve in Fig.\ref{fig2}).

\section{CP violation in the schemes with four neutrinos}
\label{CP violation in the schemes with four neutrinos}

As shown in Appendix \ref{apb},
the unitarity of the mixing matrix
implies the
``unitarity bound''
\begin{equation}
|I_{\alpha\beta}| \leq f(c_\alpha,c_\beta)
\label{Kfcc}
\end{equation}
where $f(x,y)$ is the continuous function 
\begin{equation}
f(x,y) = 
\left\{ \begin{array}{lcl} \displaystyle 
f_1 \equiv xy
& \quad \mbox{for} \quad &
2(1-x)(1-y) \ge xy
\\[3mm] \displaystyle
f_2 \equiv 2 [(x+y-1)(1-x)(1-y)]^{1/2}
& \quad \mbox{for} \quad &
2(1-x)(1-y) < xy
\end{array} \right.
\label{fxy}
\end{equation}
defined on the unit square $0 \le x \le 1$, $0 \le y \le 1$.
In Fig.\ref{fig3}
we have drawn a contour plot of the function
$f(x,y)$,
which is helpful for the determination of the maximal
allowed value for
$f(c_\alpha , c_\beta)$
when
$c_\alpha$ and/or $c_\beta$
are bounded.
The dotted line
in Fig.\ref{fig3}
is the borderline
$g(x)=2(1-x)/(2-x)$
between the regions where
$f=f_{1}$
and
$f=f_{2}$.
Note that
$f$ is continuous
along this borderline.

In order to determine the maxima of
$f(x,y)$,
the following considerations are useful
(for the details
consult Appendix \ref{apc}).
Increasing $x$ at fixed $y$,
the function $f$ increases monotonously
from $f=0$ at $x=0$,
until the straight line
$y_1(x)=2-2x$ ($ 1/2 \le x \le 1 $) 
depicted in Fig.\ref{fig3} is reached. There, the value of $f$ is
given by $f=y\sqrt{1-y}$.
After this intersection,
the function $f$ decreases monotonously to $f=0$
at $x=1$.
From the symmetry
$f(x,y)=f(y,x)$,
it follows that for fixed $x$ and
increasing $y$ the function $f$ increases monotonously
from $f=0$ at $y=0$ to $f=x\sqrt{1-x}$ when the straight line
$y_2(x)=1-x/2$ ($ 0 \le x \le 1 $)
depicted in Fig.\ref{fig3} is crossed.
After this intersection,
$f$ decreases monotonously to $f=0$
at $y=1$.

The absolute maximum of
the function $f$ (see Appendix \ref{apc})
lies at the intersection
of the lines $y_1$ and $y_2$ and is given by
$
f_{\mathrm{max}}
=
2/3\sqrt{3}
\approx
0.385
$.
Therefore,
from the unitarity of the mixing matrix
we have an absolute maximum for
$|I_{\alpha\beta}|$:
\begin{equation}
|I_{\alpha\beta}|
\leq
\frac{2}{3^{3/2}}
\approx
0.385
\;.
\label{0002}
\end{equation}

With the help of Fig.\ref{fig3},
one can see that
Eq.(\ref{Kfcc})
with the constraints (\ref{cc}) on
$c_{e}$ and $c_{\mu}$
implies that
\begin{equation}
|I_{e\mu}|
\leq
\left\{
\begin{array}{lcl} \displaystyle
f_2(a^{0}_{e},y_2(a^{0}_{e})) = 
a^{0}_{e} \left(1-a^{0}_{e}\right)^{1/2}
& \quad \mbox{for} \quad &
a^{0}_{\mu} \geq a^{0}_{e} / 2 
\;,
\\[3mm] \displaystyle
f_2(a^{0}_{e},1-a^{0}_{\mu}) =
2
\left[
\left(a^{0}_{e}-a^{0}_{\mu}\right)
\left(1-a^{0}_{e}\right)
\,
a^{0}_{\mu}
\right]^{1/2}
& \quad \mbox{for} \quad &
a^{0}_{\mu} \leq a^{0}_{e} / 2
\;.
\end{array} \right.
\label{kem1}
\end{equation}
The solid curve in Fig.\ref{fig4}
shows the limit
$ |I_{e\mu}| \leq a^{0}_{e} \, \sqrt{1-a^{0}_{e}} $
with $a^{0}_{e}$
obtained from
the 90\% CL exclusion plot of the Bugey \cite{Bugey95}
$\bar\nu_{e}\to\bar\nu_{e}$
experiment.
The dotted curve in Fig.\ref{fig4}
represents the
improvement reached with
the lower part of Eq.(\ref{kem1})
at the values of $\Delta{m}^2$
for which
$ a^{0}_{\mu} \leq a^{0}_{e} / 2 $,
with
$a^{0}_{\mu}$
obtained from the 90\% CL exclusion plots of
the CDHS \cite{CDHS84} and CCFR \cite{CCFR84}
$
\stackrel{\makebox[0pt][l]
{$\hskip-3pt\scriptscriptstyle(-)$}}{\nu_{\mu}}
\to\stackrel{\makebox[0pt][l]
{$\hskip-3pt\scriptscriptstyle(-)$}}{\nu_{\mu}}
$
experiments.

The bound represented by the solid curve in Fig.\ref{fig4}
is valid also for
$|I_{e\tau}|$,
because there is no experimental information on $c_{\tau}$.

For $|I_{\mu\tau}|$,
again by inspection of Fig.\ref{fig3},
one can see that
Eq.(\ref{Kfcc})
with the constraints (\ref{cc}) on
$c_{\mu}$
implies that
\begin{equation}
|I_{\mu\tau}|
\leq
f_2\!\left(1-a^{0}_{\mu}, y_2(1-a^0_\mu) \right) = 
\left(1-a^{0}_{\mu}\right) \sqrt{a^{0}_{\mu}}
\;.
\label{kmt1}
\end{equation}
The solid curve in Fig.\ref{fig5}
represents
the corresponding bound
obtained from the
90\% CL exclusion curves of the
CDHS \cite{CDHS84} and CCFR \cite{CCFR84}
$
\stackrel{\makebox[0pt][l]
{$\hskip-3pt\scriptscriptstyle(-)$}}{\nu_{\mu}}
\to\stackrel{\makebox[0pt][l]
{$\hskip-3pt\scriptscriptstyle(-)$}}{\nu_{\mu}}
$
experiments.
For
$ \Delta{m}^2 \lesssim 0.3 \, \mathrm{eV}^2 $
there are no experimental data
and therefore
$ |I_{\mu\tau}|_{\mathrm{max}} \approx 0.385 $
by virtue of Eq.(\ref{0002}).

Taking into account the expression
(\ref{A12})
for
$A_{\alpha;\beta}$,
in both schemes A and B
we have
(for the proof of this inequality,
see Appendix \ref{apa})
\begin{equation}
|I_{\alpha\beta}|
\leq
\frac{ 1 }{ 2 }
\sqrt{
A_{\alpha;\beta}
\left(
4 \, c_{\alpha} \, c_{\beta}
-
A_{\alpha;\beta}
\right)
}
\;.
\label{X08}
\end{equation}
Taking into account Eq.(\ref{A0}),
we obtain
\begin{equation}
|I_{\alpha\beta}|
\leq
\left\{
\begin{array}{lcl} \displaystyle
\frac{ 1 }{ 2 }
\sqrt{
A_{\alpha;\beta}^{0}
\left(
4 \, c_{\alpha} \, c_{\beta}
-
A_{\alpha;\beta}^{0}
\right)
}
& \quad \mbox{for} \quad &
A_{\alpha;\beta}^{0}
\leq
2 \, c_{\alpha} \, c_{\beta}
\;,
\\[3mm] \displaystyle
c_{\alpha} \, c_{\beta}
& \quad \mbox{for} \quad &
A_{\alpha;\beta}^{0}
\geq
2 \, c_{\alpha} \, c_{\beta}
\;.
\end{array} \right.
\label{091}
\end{equation}
For
$|I_{e\mu}|$,
with the constraints (\ref{cc}),
the inequality (\ref{091})
becomes
\begin{equation}
|I_{e\mu}|
\leq
\left\{
\begin{array}{lcl} \displaystyle
\frac{ 1 }{ 2 }
\sqrt{
A_{\mu;e}^{0}
\left(
4 \, a^{0}_{e}
-
A_{\mu;e}^{0}
\right)
}
& \quad \mbox{for} \quad &
A_{\mu;e}^{0}
\leq
2 \, a^{0}_{e}
\;,
\\[3mm] \displaystyle
a^{0}_{e}
& \quad \mbox{for} \quad &
A_{\mu;e}^{0}
\geq
2 \, a^{0}_{e}
\;.
\end{array} \right.
\label{kem2}
\end{equation}
The dash-dotted curve in Fig.\ref{fig4}
shows the limit (\ref{kem2})
obtained using the 90\% exclusion plots
of the Bugey \cite{Bugey95}
$\bar\nu_{e}\to\bar\nu_{e}$
experiment for the determination of
$a^{0}_{e}$
and
the
BNL E734
\cite{BNLE734},
BNL E776
\cite{BNLE776}
and
CCFR
\cite{CCFR96}
$\nu_{\mu}\to\nu_{e}$
experiments for the determination of
$A_{\mu;e}^{0}$.

Since the constraints (\ref{cc})
do not put an upper bound on
the possible values of
$c_{\mu}$
and
$c_{\tau}$,
in the case of
$|I_{\mu\tau}|$
the inequality (\ref{091})
becomes
\begin{equation}
|I_{\mu\tau}|
\leq
\frac{ 1 }{ 2 }
\sqrt{
A_{\mu;\tau}^{0}
\left(
4
-
A_{\mu;\tau}^{0}
\right)
}
\;.
\label{kmt2}
\end{equation}
The dotted curve in Fig.\ref{fig5}
shows the limit (\ref{kmt2})
obtained using the 90\% exclusion plot
of the
FNAL E531
\cite{FNALE531}
and
CCFR
\cite{CCFR95}
$\nu_{\mu}\to\nu_{\tau}$
experiment for the determination of
$A_{\mu;\tau}^{0}$.

The shadowed regions in Figs.\ref{fig4} and \ref{fig5}
correspond to the range (\ref{051}) of $\Delta{m}^2$
allowed at 90\% CL by the results of the LSND
and all the other SBL experiments.
From Fig.\ref{fig5}
it can be seen that,
taking into account the LSND signal,
$|I_{\mu\tau}|$
could be close to
the maximal value
$ 2/3\sqrt{3} $
allowed
by the unitarity of the mixing matrix.

\section{Three massive neutrinos}
\label{Three massive neutrinos}

It is worthwhile to have a look at LBL
neutrino oscillation experiments
neglecting some of the
present hints for neutrino oscillations.
It is possible that
not all these hints
will be substantiated in the course of time and it is
useful to check which features
are actually dependent on or independent 
from them.

In this Section we consider the minimal scenario
of mixing of three neutrinos.
We will
assume that of the two differences
of neutrino mass-squared one is
relevant for SBL
oscillations
and the other one for LBL
oscillations
(see also Refs.\cite{TA96,AR96,MN96}).
Hence,
in this section
we adopt the point of view 
that not neutrino mixing but other
reasons could explain the solar neutrino data.
With these assumptions,
there are two possible three-neutrino mass spectra:
\begin{equation}
(\mbox{\textrm{I}})
\qquad
\underbrace{
\overbrace{m_1 < m_2}^{\mathrm{LBL}}
\ll
m_3
}_{\mathrm{SBL}}
\qquad \mbox{and} \qquad 
(\mbox{\textrm{II}})
\qquad
\underbrace{
m_1
\ll
\overbrace{m_2 < m_3}^{\mathrm{LBL}}
}_{\mathrm{SBL}}
\;,
\end{equation}
In both schemes
\textrm{I} and \textrm{II},
$\Delta m^2_{31}$
is assumed to be relevant for
neutrino oscillations in SBL experiments.
In this case,
the SBL oscillation probabilities
depend on
$|U_{e3}|^2$
and
$|U_{\mu3}|^2$
in the scheme \textrm{I}
\cite{BBGK}
and
on
$|U_{e1}|^2$
and
$|U_{\mu1}|^2$
in the scheme \textrm{II}
\cite{BGKP}.
There are three regions
of these quantities
which are
allowed by the results
of disappearance experiments
(see Refs.\cite{BBGK,BGKP}):
\begin{equation}\label{regions}
\begin{array}{lll} \displaystyle
(1) \qquad &
|U_{ek}|^2 \geq 1-a^0_e
\;,
\qquad
&
|U_{{\mu}k}|^2 \leq a^0_\mu
\;,
\\[3mm] \displaystyle
(2) \qquad &
|U_{ek}|^2 \leq a^0_e
\;,
\qquad
&
|U_{\mu k}|^2 \leq a^0_\mu
\;,
\\[3mm] \displaystyle
(3) \qquad &
|U_{ek}|^2 \leq a^0_e
\;,
\qquad
&
|U_{\mu k}|^2 \geq 1-a^0_\mu
\;,
\end{array} 
\end{equation}
with
$k=3$ for the scheme \textrm{I}
and
$k=1$ for the scheme \textrm{II}\footnote{
For a comparison,
the schemes \textrm{I}, \textrm{II}
and the regions 1, 2, 3 are
called hierarchies II, I and regions A, B, C, respectively, 
in Ref.\cite{TA96}.}
(for the definition of $a^0_e$ and $a^0_\mu$,
see Eq.(\ref{a0})).

The neutrino and antineutrino
LBL oscillation probabilities
in scheme \textrm{I}
are given by
\begin{eqnarray}
&&
P^{(\mathrm{LBL},\mathrm{I})}_{\nu_\alpha\to\nu_\beta}
=
\left|
U_{\beta1}
\,
U_{\alpha1}^{*}
+
U_{\beta2}
\,
U_{\alpha2}^{*}
\,
\exp\!\left(
- i
\frac{ \Delta{m}^{2}_{21} \, L }{ 2 \, p }
\right)
\right|^2
+
|U_{{\beta}3}|^2
\,
|U_{{\alpha}3}|^2
\;,
\label{0171}
\\
&&
P^{(\mathrm{LBL},\mathrm{I})}_{\bar\nu_\alpha\to\bar\nu_\beta}
=
\left|
U_{\beta1}^{*}
\,
U_{\alpha1}
+
U_{\beta2}^{*}
\,
U_{\alpha2}
\,
\exp\!\left(
- i
\frac{ \Delta{m}^{2}_{21} \, L }{ 2 \, p }
\right)
\right|^2
+
|U_{{\beta}3}|^2
\,
|U_{{\alpha}3}|^2
\;.
\label{0172}
\end{eqnarray}
From the comparison of Eqs.(\ref{0171}) and (\ref{0172})
with Eqs.(\ref{plba1}) and (\ref{plba2}),
it is obvious that the CP-odd
asymmetries
$D^{(\mathrm{LBL},\mathrm{I})}_{\alpha;\beta}$
are given by the same formula
(\ref{073}) as in the 4-neutrino case
(with
$
I^{(\mathrm{A})}_{\alpha\beta}
\to
I^{(\mathrm{I})}_{\alpha\beta}
$).
The transition probabilities
in the scheme \textrm{II}
can be obtained from the expressions
(\ref{0171}) and (\ref{0172})
with the cyclic permutation of the indices
\begin{equation}
1 \, , \, 2 , \, 3
\rightarrow
2 \, , \, 3 , \, 1
\;.
\label{01721}
\end{equation}
Therefore,
as in the case of the schemes A and B for four neutrinos,
the bounds on the LBL oscillation probabilities
and the CP-odd asymmetries are
the same in the three neutrino schemes
\textrm{I} and \textrm{II}.

The bounds on the LBL
oscillation probabilities
$
P^{(\mathrm{LBL})}_{\stackrel{\makebox[0pt][l]
{$\hskip-3pt\scriptscriptstyle(-)$}}{\nu_{\alpha}}
\to\stackrel{\makebox[0pt][l]
{$\hskip-3pt\scriptscriptstyle(-)$}}{\nu_{\beta}}}
$
and on the CP-odd parameters
$I_{\alpha\beta}$
derived in the Appendices
for the four-neutrino schemes (\ref{AB})
are valid also in the case of mixing of three neutrinos:
the demonstrations
in the four-neutrino case A (B)
can be applied to the three neutrino case
\textrm{I} (\textrm{II})
if we put
$U_{\alpha4}=0$
($U_{\alpha1}=0$
and change the indices
$2,3,4\to1,2,3$)
for all
$\alpha=e,\mu,\tau$.

It is obvious that,
with
$ A_{\alpha;\beta} = 4 |U_{\beta 3}|^2 |U_{\alpha 3}|^2 $,
the same bounds on
$
P^{(\mathrm{LBL})}_{\stackrel{\makebox[0pt][l]
{$\hskip-3pt\scriptscriptstyle(-)$}}{\nu_{\alpha}}
\to\stackrel{\makebox[0pt][l]
{$\hskip-3pt\scriptscriptstyle(-)$}}{\nu_{\beta}}}
$
arise for $\alpha = \beta$ and $\alpha \neq \beta$
as given by Eqs.(\ref{paa})
and (\ref{pab1}).
Since in the 3-neutrino case the CP-odd asymmetries
in different oscillation channels
are connected by Eq.(\ref{CPrelations}), we have
\begin{equation}
I_{e \mu} = I_{\mu \tau} = I_{\tau e}
\;.
\label{III}
\end{equation}
A few remarks on the unitarity bound (\ref{Kfcc}) are in order. 
It is true that from the unitarity of the
$3\times 3$ mixing matrix we have $A_{\alpha;\beta} =
4(1-c_\alpha)(1-c_\beta)$ and $c_\alpha + c_\beta \geq 1$,
but,
nevertheless,
the distinction defined by Eq.(\ref{ineq})
has to be maintained.
Therefore, also the unitarity bound is upheld with the addendum that
$c_\alpha$ and $c_\beta$ can only vary within the section of the unit
square defined by $c_\alpha + c_\beta \geq 1$.
Since the point
$c_\alpha = c_\beta = 2/3$ fulfills this condition, 
the absolute maximum
$|I_{\alpha \beta}|_\mathrm{max} = 2/3\sqrt{3}$ 
of the 4-neutrino case extends its validity to three
neutrinos\footnote{This value is 4 times the maximal value of the
Jarlskog parameter 
$J$ \cite{Jarlskog} for CP violation in the Kobayashi--Maskawa matrix,
$|J|_{\mathrm{max}}=1/6\sqrt{3}$.}.

In the following we will give LBL bounds for each of the
regions (\ref{regions}),
along the lines of the previous 4-neutrino sections.

\emph{Region 1.}
With respect to SBL and LBL neutrino
oscillations,
the 3-neutrino schemes \textrm{I} and \textrm{II}
in Region 1
correspond to the
4-neutrino schemes A and B, respectively,
with the same bounds on
$
P^{(\mathrm{LBL})}_{\stackrel{\makebox[0pt][l]
{$\hskip-3pt\scriptscriptstyle(-)$}}{\nu_{e}}
\to\stackrel{\makebox[0pt][l]
{$\hskip-3pt\scriptscriptstyle(-)$}}{\nu_{e}}}
$
(Eq.(\ref{081}) and Fig.\ref{fig1}),
$
P^{(\mathrm{LBL})}_{\stackrel{\makebox[0pt][l]
{$\hskip-3pt\scriptscriptstyle(-)$}}{\nu_{\mu}}
\to\stackrel{\makebox[0pt][l]
{$\hskip-3pt\scriptscriptstyle(-)$}}{\nu_{e}}}
$
(Eqs.(\ref{042}), (\ref{0421}), (\ref{bbb1}) and Fig.\ref{fig2})
and $|I_{e\mu}|$
(Eqs.(\ref{kem1}), (\ref{kem2}) and Fig.\ref{fig4}).
Since
$ |I_{\mu\tau}| = |I_{e\mu}| $,
the stringent bounds on $|I_{e\mu}|$
given in Fig.\ref{fig4}
are valid also for $|I_{\mu\tau}|$.

For completeness,
we want to mention that there is
a change
in the upper bound for
$
P^{(\mathrm{LBL})}_{\stackrel{\makebox[0pt][l]
{$\hskip-3pt\scriptscriptstyle(-)$}}{\nu_{\mu}}
\to\stackrel{\makebox[0pt][l]
{$\hskip-3pt\scriptscriptstyle(-)$}}{\nu_{e}}}
$
in going from four to three
neutrinos:
taking into account
the inequality $ c_e + c_\mu \geq 1 $,
we have
$ c_\mu \geq 1 - \min(a^0_e,a^0_\mu) $
and Eq.(\ref{bbb1}) improves to
\begin{equation}
P^{(\mathrm{LBL})}_{\stackrel{\makebox[0pt][l]
{$\hskip-3pt\scriptscriptstyle(-)$}}{\nu_{\mu}}
\to\stackrel{\makebox[0pt][l]
{$\hskip-3pt\scriptscriptstyle(-)$}}{\nu_{e}}}
\leq
a^0_e + (1-2a^0_e) \, \min(a^0_e,a^0_\mu)
\;.
\label{newbound1}
\end{equation}
For
$ a^0_e < a^0_\mu $
this bound is slightly more stringent than that given by
Eq.(\ref{0421}),
but the improvement is negligible for
$ a^0_e \ll 1 $.

\emph{Region 2.}
Actually, this Region is excluded by the results of the
LSND experiment (see Refs. \cite{BBGK,BGKP,BGG96}).
The reason is that
(in combination with other data)
the upper bound
\begin{equation}
A_{\mu;e} \leq 4\, a^0_e a^0_\mu
\end{equation}
is too restrictive to be compatible with the LSND data.
In spite of this evidence,
let us discuss the bounds
on the LBL probabilities in
this Region.

The restrictions
$ c_e \geq 1-a^0_e $,
$ c_\mu \geq 1-a^0_\mu $
and the unitarity of the mixing matrix
imply that $ c_\tau $ is small:
$
c_\tau
=
2 - c_e - c_\mu
\leq
a^0_e + a^0_\mu
$.
From Eq.(\ref{pab1})
it follows that the probabilities of
$
\stackrel{\makebox[0pt][l]
{$\hskip-3pt\scriptscriptstyle(-)$}}{\nu_{\mu}}
\to
\stackrel{\makebox[0pt][l]
{$\hskip-3pt\scriptscriptstyle(-)$}}{\nu_{\tau}}
$
and
$
\stackrel{\makebox[0pt][l]
{$\hskip-3pt\scriptscriptstyle(-)$}}{\nu_{e}}
\to
\stackrel{\makebox[0pt][l]
{$\hskip-3pt\scriptscriptstyle(-)$}}{\nu_{\tau}}
$
transitions in LBL experiments
are confined in the range
\begin{equation}
\frac{1}{4} \, A_{\alpha;\tau}
\leq
P^{(\mathrm{LBL})}_{\stackrel{\makebox[0pt][l]
{$\hskip-3pt\scriptscriptstyle(-)$}}{\nu_{\alpha}}
\to\stackrel{\makebox[0pt][l]
{$\hskip-3pt\scriptscriptstyle(-)$}}{\nu_{\tau}}}
\leq
\frac{1}{4} \, A_{\alpha;\tau}
+ a^0_e + a^0_\mu
\qquad ( \alpha = e , \mu )
\;,
\end{equation}
whereas for the probability of
$
\stackrel{\makebox[0pt][l]
{$\hskip-3pt\scriptscriptstyle(-)$}}{\nu_{\mu}}
\to\stackrel{\makebox[0pt][l]
{$\hskip-3pt\scriptscriptstyle(-)$}}{\nu_{e}}
$
transitions we have only the lower bound
\begin{equation}
\frac{1}{4} \, A_{\mu;e}
\leq
P^{(\mathrm{LBL})}_{\stackrel{\makebox[0pt][l]
{$\hskip-3pt\scriptscriptstyle(-)$}}{\nu_{\mu}}
\to\stackrel{\makebox[0pt][l]
{$\hskip-3pt\scriptscriptstyle(-)$}}{\nu_{e}}}
\;.
\end{equation}
The inequality (\ref{pab3}) yields the
additional upper bounds
\begin{equation}
P^{(\mathrm{LBL})}_{\stackrel{\makebox[0pt][l]
{$\hskip-3pt\scriptscriptstyle(-)$}}{\nu_{\alpha}}
\to\stackrel{\makebox[0pt][l]
{$\hskip-3pt\scriptscriptstyle(-)$}}{\nu_{\tau}}}
\leq
\left( a^0_e + a^0_\mu \right)
\left( 2 - a^0_e - a^0_\mu \right)
\qquad ( \alpha = e , \mu )
\;.
\end{equation}

The bounds on CP violation can be
derived with the
methods described in the Appendices.
They are given by the
oscillation amplitude bound
\begin{equation}
|I_{e\mu}|
\leq
\frac{1}{2}
\,
\sqrt{ A_{\mu;e}^{0} \left( 4 - A_{\mu;e}^{0} \right) }
\end{equation}
and the unitarity bound
\begin{equation}
|I_{e\mu}|
\leq
f_2\!\left(1-a^{0}_{e},1-a^0_\mu\right)
=
2
\,
\sqrt{ a^0_e \, a^0_\mu \left( 1 - a^0_e - a^0_\mu \right) }
\;.
\end{equation}
Both are of similar order of magnitude and less restrictive than
the bounds in Region 1.
Since
$ |I_{\mu\tau}| = |I_{e\mu}| $,
these bounds
are valid also for $|I_{\mu\tau}|$.

\emph{Region 3.}
In this Region,
where $c_e \geq 1-a^0_e$ and
$c_\mu \leq a^0_\mu$,
the atmospheric neutrino data cannot be explained
in the framework discussed here. The reason is that
\begin{equation}
P^{(\mathrm{LBL})}_{\stackrel{\makebox[0pt][l]
{$\hskip-3pt\scriptscriptstyle(-)$}}{\nu_{\mu}}
\to\stackrel{\makebox[0pt][l]
{$\hskip-3pt\scriptscriptstyle(-)$}}{\nu_{\mu}}}
\geq (1-a^0_\mu)^2
\end{equation}
and this is incompatible \cite{BGG96}
with the atmospheric neutrino anomaly.
The LBL transition probabilities 
of muon neutrinos are confined by
\begin{equation}
\frac{1}{4} \, A_{\mu;\beta}
\leq
P^{(\mathrm{LBL})}_{\stackrel{\makebox[0pt][l]
{$\hskip-3pt\scriptscriptstyle(-)$}}{\nu_{\mu}}
\to\stackrel{\makebox[0pt][l]
{$\hskip-3pt\scriptscriptstyle(-)$}}{\nu_{\beta}}}
\leq
\frac{1}{4} \, A_{\mu;\beta} + a^0_\mu
\qquad
( \beta = e , \tau )
\;,
\end{equation}
whereas for
$
\stackrel{\makebox[0pt][l]
{$\hskip-3pt\scriptscriptstyle(-)$}}{\nu_{e}}
\to\stackrel{\makebox[0pt][l]
{$\hskip-3pt\scriptscriptstyle(-)$}}{\nu_{\tau}}
$
transitions
there is only the
lower bound
\begin{equation}
\frac{1}{4} \, A_{e;\tau}
\leq
P^{(\mathrm{LBL})}_{\stackrel{\makebox[0pt][l]
{$\hskip-3pt\scriptscriptstyle(-)$}}{\nu_{e}}
\to\stackrel{\makebox[0pt][l]
{$\hskip-3pt\scriptscriptstyle(-)$}}{\nu_{\tau}}}
\;.
\end{equation}
The inequality (\ref{pab2}),
which is a consequence of
probability conservation,
leads to
\begin{equation}
P^{(\mathrm{LBL})}_{\stackrel{\makebox[0pt][l]
{$\hskip-3pt\scriptscriptstyle(-)$}}{\nu_{\mu}}
\to\stackrel{\makebox[0pt][l]
{$\hskip-3pt\scriptscriptstyle(-)$}}{\nu_{\beta}}}
\leq
a^0_\mu \left( 2 - a^0_\mu \right) 
\qquad
( \beta = e , \tau )
\;.
\label{oldbound3}
\end{equation}
Furthermore,
taking into account
the inequality $ c_e + c_\mu \geq 1 $,
we have
$ c_e \geq 1 - \min(a^0_e,a^0_\mu) $
and Eq.(\ref{bbb1}) improves to
\begin{equation}
P^{(\mathrm{LBL})}_{\stackrel{\makebox[0pt][l]
{$\hskip-3pt\scriptscriptstyle(-)$}}{\nu_{\mu}}
\to\stackrel{\makebox[0pt][l]
{$\hskip-3pt\scriptscriptstyle(-)$}}{\nu_{e}}}
\leq
a^0_\mu + (1-2a^0_\mu) \, \min(a^0_e,a^0_\mu)
\;.
\label{newbound3}
\end{equation}
For
$ a^0_e \ll a^0_\mu \ll 1 $
this bound is about half of that given by
Eq.(\ref{oldbound3}).

Finally, our methods for obtaining bounds on CP violation yield
\begin{equation}
|I_{e\mu}|
\leq
\frac{1}{2}
\,
\sqrt{ A_{\mu;e}^{0} \left( 4 \, a^0_\mu - A_{\mu;e}^{0} \right) }
\label{I31}
\end{equation}
and
\begin{equation}
| I_{e \mu}| \leq \left\{
\begin{array}{lcl} \displaystyle
a^0_\mu
\,
\left( 1 - a^0_\mu \right)^{1/2}
& \quad \mbox{for} \quad & 
a^0_e \geq a^0_\mu / 2
\;,
\\[1mm] \displaystyle
2
\,
\left[ \left( a^0_\mu - a^0_e \right)
\left( 1 - a^0_\mu \right) a^0_e \right]^{1/2}
& \quad \mbox{for} \quad & 
a^0_e \leq a^0_\mu / 2
\;.
\end{array} \right.
\label{I32}
\end{equation}
The oscillation amplitude bound is more stringent in this case.
From Eq.(\ref{III}) it follows that
the bounds (\ref{I31}) and (\ref{I32})
are valid also for the parameter
$|I_{\mu\tau}|$
that characterizes the CP-odd asymmetry in the
$\nu_\mu\to\nu_\tau$
channel.

The differences in the bounds on the LBL
probabilities are marked and could thus serve
to distinguish between the three
different Regions in the three neutrino case. Of course, in the
experiments discussed here the four neutrino 
case (schemes A and B) is indistinguishable from the 
three neutrino case with Region 1.
The above-mentioned distinctions could also serve as a cross-check
for present hints of neutrino oscillations.

\section{Conclusions}
\label{Conclusions}

At present there are three experimental indications
in favour of neutrino
oscillations which correspond to three different scales of
neutrino mass-squared differences:
the solar neutrino deficit, the atmospheric neutrino
anomaly and the result of the LSND experiment.
These indications
and the negative results of numerous short-baseline neutrino
experiments can be accommodated in two
schemes (A and B) with mixing of four massive neutrinos
\cite{BGG96}.
In this paper
we have presented a detailed study of the
predictions of the schemes A and B for
long-baseline experiments.
Using only the results
of existing experiments,
we have
obtained bounds on probabilities of different transitions in
long-baseline experiments.

The schemes A and B give completely different predictions for
neutrinoless double beta decay and for neutrino mass effects in
experiments for neutrino mass measurements by the tritium method
\cite{BGG96}.
They lead, however,
to the same bounds on long-baseline oscillation probabilities.
In addition, all the bounds that we have derived
apply for neutrinos as well as antineutrinos.

We have shown that the results of the
short-baseline reactor experiment put
rather severe bounds on the probability
$1-P^{(\mathrm{LBL})}_{\bar\nu_e \to \bar\nu_e}$
of $\bar\nu_e$ transitions
into all possible other states
in long-baseline experiments.
If the $\Delta{m}^2$ relevant in short-baseline oscillations
is bigger than about
$ 3 \, \mathrm{eV}^2 $,
the bound on
$1-P^{(\mathrm{LBL})}_{\bar\nu_e \to \bar\nu_e}$
is slightly higher
than the sensitivity of the CHOOZ experiment,
allowing some possibility to
reveal neutrino oscillations in this channel.
However,
the results of the
LSND experiment favour the range
$ 0.3 \lesssim \Delta{m}^2 \lesssim 2.2 \, \mathrm{eV}^2 $.
We have
shown that in this range the
upper bound for the quantity 
$1-P^{(\mathrm{LBL})}_{\bar\nu_e\to\bar\nu_e}$
lies between
$10^{-2}$ and $5\times10^{-2}$
(see Fig.\ref{fig1}) and thus below the sensitivity
of CHOOZ and Palo Verde.
On the other hand,
there is no restriction on
$
P^{(\mathrm{LBL})}_{\stackrel{\makebox[0pt][l]
{$\hskip-3pt\scriptscriptstyle(-)$}}{\nu_{\mu}}
\to\stackrel{\makebox[0pt][l]
{$\hskip-3pt\scriptscriptstyle(-)$}}{\nu_{\mu}}}
$. 

The probability
$
P^{(\mathrm{LBL})}_{\stackrel{\makebox[0pt][l]
{$\hskip-3pt\scriptscriptstyle(-)$}}{\nu_{\mu}}
\to\stackrel{\makebox[0pt][l]
{$\hskip-3pt\scriptscriptstyle(-)$}}{\nu_{e}}}
$
of
$
\stackrel{\makebox[0pt][l]
{$\hskip-3pt\scriptscriptstyle(-)$}}{\nu_{\mu}}
\to\stackrel{\makebox[0pt][l]
{$\hskip-3pt\scriptscriptstyle(-)$}}{\nu_{e}}
$
transitions
is severely
constrained in the schemes A and B
by the results of short-baseline
reactor and accelerator experiments
(see Fig.\ref{fig2}). 
The sensitivity of MINOS and ICARUS
is well below the upper bound for
$
P^{(\mathrm{LBL})}_{\stackrel{\makebox[0pt][l]
{$\hskip-3pt\scriptscriptstyle(-)$}}{\nu_{\mu}}
\to\stackrel{\makebox[0pt][l]
{$\hskip-3pt\scriptscriptstyle(-)$}}{\nu_{e}}}
$,
whereas the sensitivity of
the KEK-SK experiment seems to be insufficient.
There is also an
upper bound on long-baseline
$
\stackrel{\makebox[0pt][l]
{$\hskip-3pt\scriptscriptstyle(-)$}}{\nu_{e}}
\to\stackrel{\makebox[0pt][l]
{$\hskip-3pt\scriptscriptstyle(-)$}}{\nu_{\tau}}
$
oscillations
(solid curve in Fig.\ref{fig2}),
which is nearly as tight as the one 
for
$
\stackrel{\makebox[0pt][l]
{$\hskip-3pt\scriptscriptstyle(-)$}}{\nu_{\mu}}
\to\stackrel{\makebox[0pt][l]
{$\hskip-3pt\scriptscriptstyle(-)$}}{\nu_{e}}
$
transitions.
On the other hand,
the long-baseline
$
\stackrel{\makebox[0pt][l]
{$\hskip-3pt\scriptscriptstyle(-)$}}{\nu_{\mu}}
\to\stackrel{\makebox[0pt][l]
{$\hskip-3pt\scriptscriptstyle(-)$}}{\nu_{\tau}}
$
channel is unconstrained.

We have obtained bounds on LBL transition probabilities
in the case of the neutrino mass spectra (\ref{AB}),
which are implied by the results
of the solar, atmospheric and
LSND experiments.
If the LSND data are not confirmed by future experiments,
but nevertheless there is a mass (or masses) approximately
equal to 1 eV
providing an explanation for
the hot dark matter problem,
then the
neutrino mass spectrum can be different from
the spectra A and B in
Eq.(\ref{AB}).
The natural neutrino mass spectrum in this case is hierarchical
and in this case the bounds
that we have obtained in this
paper are not valid.

In the framework of neutrino mixing in schemes A and B,
we have also
derived constraints on the
parameters $I_{\alpha \beta}$
that characterize
the CP-odd neutrino--antineutrino
asymmetries in long-baseline experiments.
We have developed methods for deriving upper bounds on the
parameters $I_{\alpha \beta}$
from the data of
short-baseline experiments,
which can be applied to different
schemes.
We have shown that CP violation in the
$\nu_\mu\rightarrow\nu_e$
channel is bounded by
$| I_{e\mu} | \lesssim 10^{-2}$.
A similar suppression of CP-odd effect takes place in
$\nu_e\rightarrow\nu_\tau$
long-baseline neutrino oscillations.
On the other hand,
sizable CP violation can be expected in
$\nu_\mu \rightarrow \nu_\tau$
oscillations.
The CP-odd asymmetry in this channel
could be close to its maximally allowed
value
$|I_{\mu\tau}|_{\mathrm{max}} \approx 0.385$, 
resulting from unitarity of the mixing matrix.

Summarizing,
we would like to emphasize that 
the results of all neutrino oscillation experiments
lead to severe
constraints on the
probabilities for
the probabilities for
$\bar\nu_e$
disappearance and
$
\stackrel{\makebox[0pt][l]
{$\hskip-3pt\scriptscriptstyle(-)$}}{\nu_{\mu}}
\to\stackrel{\makebox[0pt][l]
{$\hskip-3pt\scriptscriptstyle(-)$}}{\nu_{e}}
$
and
$
\stackrel{\makebox[0pt][l]
{$\hskip-3pt\scriptscriptstyle(-)$}}{\nu_{e}}
\to\stackrel{\makebox[0pt][l]
{$\hskip-3pt\scriptscriptstyle(-)$}}{\nu_{\tau}}
$
appearance
in long-baseline experiments.
Nevertheless, even these channels
are within the planned
sensitivity of the MINOS and ICARUS experiments.
The channels
$
\stackrel{\makebox[0pt][l]
{$\hskip-3pt\scriptscriptstyle(-)$}}{\nu_{\mu}}
\to\stackrel{\makebox[0pt][l]
{$\hskip-3pt\scriptscriptstyle(-)$}}{\nu_{\tau}}
$
and
$
\stackrel{\makebox[0pt][l]
{$\hskip-3pt\scriptscriptstyle(-)$}}{\nu_{\mu}}
\to\stackrel{\makebox[0pt][l]
{$\hskip-3pt\scriptscriptstyle(-)$}}{\nu_{\mu}}
$
are not constrained at all.
Therefore,
from the point of view of the present investigation, 
long-baseline muon neutrino beams provide promising
facilities for the observation of neutrino oscillations.

Let us stress that we have considered
here vacuum LBL oscillations.
In conclusion, we want to present some
remarks about the effects of matter.
In the 4-neutrino schemes under
consideration active neutrinos can transform into sterile
states.
Therefore, not only charged current but also neutral current
interactions must be taken into account \cite{MSW} and the effective
Hamiltonian in the flavour representation is given by
\begin{equation}\label{Heff}
H_{\mathrm{eff}} = \frac{1}{2p} \left( U \hat{M}^2 U^\dagger + 
\mathrm{diag}\, (a_{CC}, 0, 0, a_{NC}) \right)
\end{equation}
with
\begin{eqnarray}
&&
a_{CC}
=
2\sqrt{2} \, G_F \, N_e \, p
\simeq
8 \times 10^{-5} \, \mathrm{eV}^2
\left( \frac{ \rho }{ 1 \, \mathrm{g} \, \mathrm{cm}^{-3} } \right)
\left( \frac{ p }{ 1 \, \mathrm{GeV} } \right)
\;,
\label{acc}
\\
&&
a_{NC}
=
\sqrt{2} \, G_F \, N_n \, p
\simeq
\frac{1}{2} \, a_{CC}
\label{anc}
\end{eqnarray}
where $\hat{M}^2$ is the diagonal matrix of the squared masses,
$G_F$ is the Fermi constant,
$N_e$ and $N_n$ are the electron and neutron number
density\footnote{$
N_e \simeq N_n \simeq
\frac{N_{\mathrm{A}}}{2}
\frac{\rho}{1\mathrm{g}}
$,
where
$N_{\mathrm{A}}$
is the Avogadro number.
},
respectively,
and
$\rho$ is the density of matter,
which in the Earth's crust is of the order of
$ 3 \, \mathrm{g} \, \mathrm{cm}^{-3} $.
The parameters $a_{CC}$ and $a_{NC}$ are small with respect to the
$\delta m^2$ relevant for LBL oscillations. Nevertheless, their
contributions for some transitions can be important (see \cite{MN96}
for the case of three neutrinos).

In order to take into account
the matter effects in the oscillation probabilities we
must diagonalize
$H_{\mathrm{eff}}$\footnote{
Here we consider the simplifying
approximation of constant $N_e$ and $N_n$.
}.
In the case of the schemes (\ref{AB}) under consideration,
the relation $a_{CC} \ll \Delta m^2 \sim 1 \mathrm{eV}^2$
is valid and thus,
up to terms of the order $a_{CC}/\Delta m^2$,
we have
\begin{equation}
H_{\mathrm{eff}} = U' \frac{\hat{\epsilon}}{2p} U'^\dagger
\end{equation}
where $\hat{\epsilon} = \mathrm{diag}\, (\epsilon_1, \ldots ,
\epsilon_4)$, $U' = UR$ and $R$ has the block structure
\begin{equation}
R = \left( \begin{array}{cc}    
           R' & 0 \\ 0 & R'' \end{array} \right).
\end{equation}
The matrices $R'$ and $R''$ are 2$\times$2 unitary matrices. Now
instead of Eq.(3.7) we obtain for the LBL oscillation probabilities
\begin{equation}
P^{(\mathrm{LBL})}_{\nu_\alpha\to\nu_\beta} =
\left| \sum_{j=1,2} U'_{\beta j}U'^*_{\alpha j}
\exp{\left(-i \frac{\epsilon_j}{2p} \, L\right)} \right|^2 +
\left| \sum_{k=3,4} U'_{\beta k}U'^*_{\alpha k}
\exp{\left(-i \frac{\epsilon_k}{2p} \, L\right)} \right|^2\, .
\end{equation}

From the Cauchy--Schwarz inequality it is easy to see that in both
schemes A and B the upper bounds (\ref{paa}), (\ref{bbb})
and therefore (\ref{bbb1})
(represented by the short-dashed curve in Fig.\ref{fig2})
are valid in
the case of matter apart from corrections of the order of
$a_{CC}/\Delta m^2$.
Other bounds could be modified by the matter effect.

However,
the upper bound (\ref{081}) applied to reactor LBL
experiments is practically not modified by matter effects.
In fact,
the relevant parameter $\epsilon_4 - \epsilon_3 \sim a_{CC}L/2p
\approx 0.5 \times 10^{-6} \, (L/1\, \mathrm{m})$, for $\rho = 3 \,
\mathrm{g}\, \mathrm{cm}^{-3}$, is around $10^{-4}$ for CHOOZ and Palo
Verde, whereas in the case of MINOS and ICARUS it is not small
($a_{CC}L/2p \approx 0.4$). For the KEK-SK experiment, matter
corrections are still modest because the baseline is roughly a factor
of three smaller than the baseline of MINOS and ICARUS.
Due to matter effects the bounds (\ref{042}) and (\ref{0421}) on the
$\nu_\mu\to\nu_e$ transition probability depicted in Fig.\ref{fig2}
become less
restrictive, in particular for the MINOS and ICARUS experiments.
Note
that also because matter induces corrections of the order of $10^{-2}$
the lower bound obtained in the vacuum case from the results of the
LSND experiment
(the left borderline of the darkly shadowed area in Fig.\ref{fig2})
disappears
and the lightly shadowed
region in Fig.\ref{fig2} is allowed by the LSND experiment
in the case of LBL oscillations in matter.
A detailed discussion of
matter effects in the case of mixing of four massive neutrinos will be
published elsewhere \cite{BGG97}.

For antineutrinos the relevant effective Hamiltonian is obtained from
Eq.(\ref{Heff}) by taking $U^*$ instead of $U$ and putting a minus in
front of $a_{CC}$ and $a_{NC}$. Therefore, matter effects can mimic
CP-odd asymmetries and at least at the order of $10^{-2}$ both effects
are entangled.
In the framework of the schemes
with mixing of three neutrinos
the effects of matter in long-baseline experiments
were considered recently in Refs.\cite{TA96,AR96,MN96}.
It was shown that
the contribution of matter to the CP-odd asymmetries
is small for all transitions,
except for
$\nu_\mu\to\nu_e$
transitions in the Region 2 of the scheme \textrm{I}
\cite{MN96},
which is not compatible with the LSND result
\cite{BBGK}.

\acknowledgments

This work was done while one of
authors (S.M.B.) was  
Lady Davis visiting professor at the Technion.
This author would like to thank
the Physics Department of Technion 
for its hospitality.

\newpage
\appendix

\section{Derivation of the oscillation amplitude bound}
\label{apa}

In this Appendix we discuss
the derivation of the ``oscillation 
amplitude  bound''.
The starting point is the quantity
\begin{equation}
I_{\alpha\beta}
=
4
\,
\mbox{Im}\!\left[
U_{\alpha1} \, U_{\beta1}^* \, U_{\alpha2}^* \, U_{\beta2}
\right]
\qquad (\alpha \neq \beta)
\;,
\end{equation}
which determines the CP-odd asymmetry
in the case of four massive neutrinos. 

It is obvious that $I_{\alpha\beta}$ is invariant under the phase 
transformation
\begin{equation}
U_{\alpha j} \rightarrow \mbox{e}^{i \gamma_j} U_{\alpha j}\;,
\qquad
U_{\beta j}  \rightarrow \mbox{e}^{i \gamma_j}  U_{\beta j}
\;, 
\end{equation}
where the $\gamma_j$ are arbitrary phases.
Thus the elements $U_{\alpha j}$
can be taken to be real.
Taking into account
the definitions (\ref{dca}) and (\ref{dcb})
valid in the schemes A and B of Eq.(\ref{AB}),
respectively,
we can write
\begin{equation}
U_{\alpha j} = \sqrt{c_\alpha} \, e^{(1)}_j
\;,
\qquad
\mbox{with}
\quad
\left\{
\begin{array}{rcl} \displaystyle
j=1,2
& \quad & \displaystyle
\mbox{in the scheme A,}
\\ \displaystyle
j=3,4
& \quad & \displaystyle
\mbox{in the scheme B,}
\end{array}
\right.
\label{Ual}
\end{equation}
and the orthonormal basis
\begin{equation}\label{ON}
e^{(1)}(\theta) = (\cos \theta, \sin \theta)
\;,
\qquad
e^{(2)}(\theta) = (-\sin \theta, \cos \theta)
\;.
\end{equation}
We expand $U_{\beta j}$
(with
$j=1,2$
in the scheme A
and
$j=3,4$
in the scheme B)
with respect to this
basis as
\begin{equation}\label{Ube}
U_{\beta j}
=
\sqrt{c_\beta} \sum_{\rho=1,2} p_\rho \, e^{(\rho)}_j
\;, 
\end{equation}
where
$p_1$ and $p_2$
are complex coefficients
such that 
\begin{equation}\label{norm}
\sum_{\rho=1,2} |p_\rho|^2 = 1
\;.
\end{equation}
With the help of
Eqs.(\ref{Ual})--(\ref{norm})
we easily find
\begin{equation}\label{Kp}
I_{\alpha\beta}
=
2 \, c_\alpha \, c_\beta \, \sin 2\theta \, 
\mbox{Im}(p_1^* p_2) =
2 \, c_\alpha \, c_\beta
\, |p_1| \, \sqrt{1-|p_1|^2}
\, \sin 2\theta \, \sin \delta
\;,
\end{equation}
where $\delta$ is the phase of $p_1^* p_2$.

The parameter $|p_1|$ is connected to the oscillation amplitude
$A_{\alpha;\beta}$ and the parameters $c_\alpha$, $c_\beta$.
In fact,
from Eqs.(\ref{Ual}) and (\ref{Ube}) we have
\begin{equation}
A_{\alpha;\beta}
=
\left\{
\begin{array}{rcl} \displaystyle
4 \left| \sum_{j=1,2} U_{\alpha j} U_{\beta j}^* \right|^2
=
4 \, c_\alpha \, c_\beta \, |p_1|^2
& \quad & \displaystyle
\mbox{in the scheme A}
\;,
\\ \displaystyle
4 \left| \sum_{j=3,4} U_{\alpha j} U_{\beta j}^* \right|^2
=
4 \, c_\alpha \, c_\beta \, |p_1|^2
& \quad & \displaystyle
\mbox{in the scheme B}
\;.
\end{array}
\right.
\end{equation}
Hence,
in both schemes A and B,
we have
$ |p_1| = \sqrt{ A_{\alpha;\beta} / 4 c_\alpha c_\beta } $.
Inserting this value in
Eq.(\ref{Kp}),
we obtain
\begin{equation}
I_{\alpha\beta} =
\frac{1}{2}
\,
\sqrt{
A_{\alpha;\beta}
\left(
4 \, c_{\alpha} \, c_{\beta}
-
A_{\alpha;\beta}
\right)
}
\, \sin 2\theta \, \sin \delta
\end{equation}
and thus we arrive at the ``oscillation amplitude bound''
\begin{equation}
|I_{\alpha\beta}|
\leq
\frac{1}{2} \,
\sqrt{
A_{\alpha;\beta} 
\left( 
4 \, c_{\alpha} \, c_{\beta} - A_{\alpha;\beta}
\right)
}
\;.
\label{oab}
\end{equation}
Let us stress that this derivation is based only on the obvious
inequality
\begin{equation}
| \sin 2\theta \sin \delta | \leq 1
\;.
\end{equation}
Since $c_\alpha$, $c_\beta$ and
$A_{\alpha;\beta}$ do not restrict $\theta$ and $\delta$,
the bound
(\ref{oab}) is the optimal one.

\section{Derivation of the unitarity bound}
\label{apb}

Up to now we did not use
the unitarity of the mixing matrix.
Taking this fact
into account will allow us to obtain an upper bound on
$|I_{\alpha\beta}|$
depending solely on $c_\alpha$ and $c_\beta$.

The unitarity of the mixing matrix tells us that
\begin{equation}
\sum_{j=1,2} U_{\alpha j} U_{\beta j}^*
=
-
\sum_{j=3,4} U_{\alpha j} U_{\beta j}^*
\;.
\end{equation}
This relation allows to write the
oscillation amplitude
$A_{\alpha;\beta}$
in the two forms of Eq.(\ref{A12}).
Using the Cauchy--Schwarz inequality,
one can see that
\begin{equation}
A_{\alpha;\beta}
=
4
\left| \sum_{j=3,4} U_{\alpha j} U_{\beta j}^* \right|^2
\leq
4
\left( \sum_{j=3,4} |U_{\alpha j}|^2 \right)
\left( \sum_{j=3,4} |U_{\beta j}|^2 \right)
=
4 \, (1-c_\alpha) \, (1-c_\beta)
\;,
\label{csa}
\end{equation}
in the scheme A,
and
\begin{equation}
A_{\alpha;\beta}
=
4
\left| \sum_{j=1,2} U_{\alpha j} U_{\beta j}^* \right|^2
\leq
4
\left( \sum_{j=1,2} |U_{\alpha j}|^2 \right)
\left( \sum_{j=1,2} |U_{\beta j}|^2 \right)
=
4 \, (1-c_\alpha) \, (1-c_\beta)
\;,
\label{csb}
\end{equation}
in the scheme B.
Hence,
in both schemes A and B
the oscillation amplitude
$A_{\alpha;\beta}$
is bounded by
\begin{equation}
A_{\alpha;\beta}
\leq
4 \, (1-c_\alpha) \, (1-c_\beta)
\;.
\label{cs}
\end{equation}

The right-hand side of the inequality
(\ref{oab}),
as a function of $A_{\alpha;\beta}$,
reaches its maximum,
$ c_\alpha c_\beta $,
at
\begin{equation}
(A_{\alpha;\beta})_0
=
2 \, c_\alpha \, c_\beta
\;. 
\end{equation}
Consequently, if the condition
\begin{equation}\label{ineq}
2 \, (1-c_\alpha) \, (1-c_\beta)
\geq
c_\alpha \, c_\beta
\end{equation}
is satisfied,
the upper bound (\ref{cs}) on
$A_{\alpha;\beta}$
is larger then
$(A_{\alpha;\beta})_0$.
In this case we have
\begin{equation}
|I_{\alpha\beta}| \leq c_\alpha \, c_\beta
\;.
\end{equation}
If the condition (\ref{ineq}) is not fulfilled,
the upper bound
(\ref{cs}) is smaller than $(A_{\alpha;\beta})_0$
and has to be
inserted for $A_{\alpha;\beta}$
into Eq.(\ref{oab}),
leading to
\begin{equation}
|I_{\alpha\beta}| \leq
2 \,
\sqrt{
(c_\alpha + c_\beta -1) \, (1-c_\alpha) \, (1-c_\beta)
}
\;.
\end{equation}
Thus, we arrive at the ``unitarity bound''
\begin{equation}
|I_{\alpha\beta}| \leq f(c_\alpha , c_\beta)
\;,
\end{equation}
with the function 
\begin{equation}\label{f}
f(x,y) = 
\left\{ \begin{array}{lcl} \displaystyle 
f_1 \equiv xy
& \quad \mbox{for} \quad &
2(1-x)(1-y) \ge xy
\;,
\\[3mm] \displaystyle
f_2 \equiv 2 [(x+y-1)(1-x)(1-y)]^{1/2}
& \quad \mbox{for} \quad &
2(1-x)(1-y) < xy
\;,
\end{array} \right.
\end{equation}
defined on the unit square $0 \le x \le 1$, $0 \le y \le 1$.
The
function
\begin{equation}
g(x) = \frac{2\,(1-x)}{2-x}
\end{equation}
represents the borderline separating the two regions in the
definition of the function (\ref{f}).
It is clear from our
derivation (and also easy to check)
that $f$
is continuous along this borderline.

\section{Discussion of the function $\lowercase{f}$}
\label{apc}

Since we do not have
definite experimental values of
$c_\alpha$
and
$c_\beta$,
but only bounds on these quantities
(see Eq.(\ref{cc})),
which
define allowed rectangles in the square
$0 \leq c_\alpha \leq 1$,
$0 \leq c_\beta \leq 1$,
we are interested in the behaviour of $f$
in order to evaluate the
unitarity bound.

From the partial derivative of $f$
in the region $y \ge g(x)$, 
\begin{equation}\label{part}
\frac{\partial f}{\partial x} = \frac{\partial f_2}{\partial x}
\propto (2-2x-y)
\;,
\end{equation}
one can see that,
at fixed $y$,
the function $f$ increases monotonously
from $x=0$ to the point $x=1-y/2$,
where the partial derivative in Eq.(\ref{part}) is zero.
The points
$x=1-y/2$ lie on the straight line $y_1(x)=2-2x$.
In the range
$1-y/2 \leq x \leq 1$
the function $f$ decreases monotonously.
Taking
into account the symmetry $f(x,y)=f(y,x)$,
we see that at fixed $x$ the function $f$
increases monotonously from $y=0$ to the point $y=1-x/2$,
where the
partial derivative of $f$ with respect to $y$ is zero.
These points
lie on the straight line $y_2(x)=1-x/2$.
Beyond this line $f$
decreases monotonously. Note that both straight lines lie in the range
of $f_2$.

Figure \ref{fig3} shows a contour plot of
the function $f(x,y)$,
together with the lines
$y_1$ and $y_2$ which intersect at the point
\begin{equation}\label{point}
x = y = \frac{2}{3}
\;.
\end{equation}
At this point both partial derivatives of $f$
are equal to zero and 
therefore the point (\ref{point})
corresponds to the absolute maximum of $f$,
given by
\begin{equation}
f_{\mbox{\scriptsize max}}
=
f_2\!\left(\frac{2}{3},\frac{2}{3}\right)
=
\frac{2}{3^{3/2}} 
\approx
0.385
\;.
\end{equation}
This number constitutes the absolute
upper bound for $|I_{\alpha\beta}|$.

\begin{figure}[h]
\refstepcounter{figure}
\label{fig1}
FIG.\ref{fig1}.
Upper bound for
the transition probability of
LBL reactor $\bar\nu_e$'s
into all possible states,
$1-P^{(\mathrm{LBL})}_{\nu_e\to\nu_e}$,
for the SBL
$\Delta{m}^2$
in the range
$
10^{-1} \, \mathrm{eV}^2
\leq \Delta{m}^2 \leq
10^{3} \, \mathrm{eV}^2
$.
The solid curve is obtained 
from the 90\% CL exclusion plot of the Bugey
$\bar\nu_e\to\bar\nu_e$ experiment
(see Eq.(\ref{081})).
The dash-dotted and dash-dot-dotted vertical lines
depict, respectively, the expected sensitivities
of the CHOOZ and Palo Verde
LBL reactor neutrino experiments.
The shadowed region
corresponds to the range of $\Delta{m}^2$
allowed at 90\% CL by the results of the LSND experiment,
taking into account the results of
all the other SBL experiments
(see Eq.(\ref{051})).
\end{figure}

\begin{figure}[h]
\refstepcounter{figure}
\label{fig2}
FIG.\ref{fig2}.
Upper bound for
the probability 
of
$\nu_{\mu}\to\nu_{e}$
and
$\bar\nu_\mu\to\bar\nu_e$
transitions in LBL experiments.
The solid curve is obtained 
only from the 90\% CL exclusion plot of the Bugey
$\bar\nu_e\to\bar\nu_e$ experiment
(see Eq.(\ref{0421})),
whereas
for the long-dashed curve
the 90\% CL exclusion plots of the
BNL E734,
BNL E776 and
CCFR
$\nu_\mu\to\nu_e$
and
$\bar\nu_\mu\to\bar\nu_e$
experiments
(see Eq.(\ref{042}))
and
for the short-dashed curve
the 90\% CL exclusion plots
of the CDHS \cite{CDHS84} and CCFR \cite{CCFR84}
$\bar\nu_{\mu}\to\bar\nu_{\mu}$
experiments
(see Eq.(\ref{bbb1}))
have been used in addition.
The dotted, dash-dotted and dash-dot-dotted
vertical lines represent,
respectively,
the expected sensitivities
of the LBL accelerator neutrino experiments
KEK--SK,
MINOS
and
ICARUS.
The darkly shadowed region is allowed
by the results of the LSND experiment,
taking into account the results of
all the other SBL experiments,
in the case of LBL oscillations in vacuum.
The two horizontal borderlines correspond
to the limits (\ref{051}) for $\Delta m^2$
and the left
borderline corresponds to the lower bound
in Eq.(\ref{pab12}).
The lightly shadowed region
is allowed
by the results of the LSND and
all the other SBL experiments
in the case of LBL oscillations in matter.
The solid curve constitutes also an upper bound for
the probability
of
$\nu_{e}\to\nu_{\tau}$
and
$\bar\nu_{e}\to\bar\nu_{\tau}$
transitions.
\end{figure}

\begin{figure}[h]
\refstepcounter{figure}
\label{fig3}
FIG.\ref{fig3}.
Contour plot of the function
$f(x,y)$ given in Eq.(\ref{fxy}).
The dotted line
is the borderline
$g(x)=2(1-x)/(2-x)$
between the regions where
$f=f_{1}$
and
$f=f_{2}$.
The two solid lines represent the functions
$y_1(x)=2-2x$
and 
$y_2(x)=1-x/2$.
\end{figure}

\begin{figure}[h]
\refstepcounter{figure}
\label{fig4}
FIG.\ref{fig4}.
Upper bound for
the parameter
$|I_{e\mu}|$
which characterizes the CP-odd asymmetry
in the $\nu_{\mu}\to\nu_{e}$ channel
for the SBL parameter
$\Delta{m}^2$
in the range
$
10^{-1} \, \mathrm{eV}^2
\leq \Delta{m}^2 \leq
10^{3} \, \mathrm{eV}^2
$.
The solid curve
represents the upper function
in Eq.(\ref{kem1})
and is obtained 
from the 90\% CL exclusion plot of the Bugey
$\bar\nu_e\to\bar\nu_e$ experiment.
The dotted curve improves the solid curve
where $a^0_\mu \leq a^0_e/2$
(the lower function in Eq.(\ref{kem1})).
It is obtained
from the 90\% CL exclusion plots of the Bugey
$\bar\nu_e\to\bar\nu_e$ experiment
and the CDHS and CCFR
$\nu_{\mu}\to\nu_{\mu}$
and
$\bar\nu_{\mu}\to\bar\nu_{\mu}$
experiments.
The dash-dotted curve is obtained 
from the 90\% CL exclusion plots of the Bugey
$\bar\nu_e\to\bar\nu_e$ experiment
and the
BNL E734,
BNL E776 and
CCFR
$\nu_\mu\to\nu_e$
and
$\bar\nu_\mu\to\bar\nu_e$
experiments
(see the upper function in  Eq.(\ref{kem2})).
The shadowed region
corresponds to the range (\ref{051}) of $\Delta{m}^2$
allowed at 90\% CL by the results of the LSND experiment.
The solid curve represents also an upper bound for
$|I_{e\tau}|$.
\end{figure}

\begin{figure}[h]
\refstepcounter{figure}
\label{fig5}
FIG.\ref{fig5}.
Upper bound for
the parameter $|I_{\mu\tau}|$
which characterizes the CP-odd asymmetry
in the $\nu_{\mu}\to\nu_{\tau}$ channel.
The solid curve is obtained 
from the 90\% CL exclusion plots of the CDHS and CCFR
$\nu_{\mu}\to\nu_{\mu}$
and
$\bar\nu_{\mu}\to\bar\nu_{\mu}$
experiments
(see Eq.(\ref{kmt1})).
The dotted curve is obtained 
from the 90\% CL exclusion plots of the FNAL E531 and CCFR
$\nu_\mu\to\nu_\tau$
experiments
(see Eq.(\ref{kmt2})).
The shadowed region
corresponds to the range (\ref{051}) of $\Delta{m}^2$
allowed at 90\% CL by the results of the LSND experiment.
\end{figure}

\newpage

\begin{minipage}[p]{0.95\textwidth}
\begin{center}
\mbox{\epsfig{file=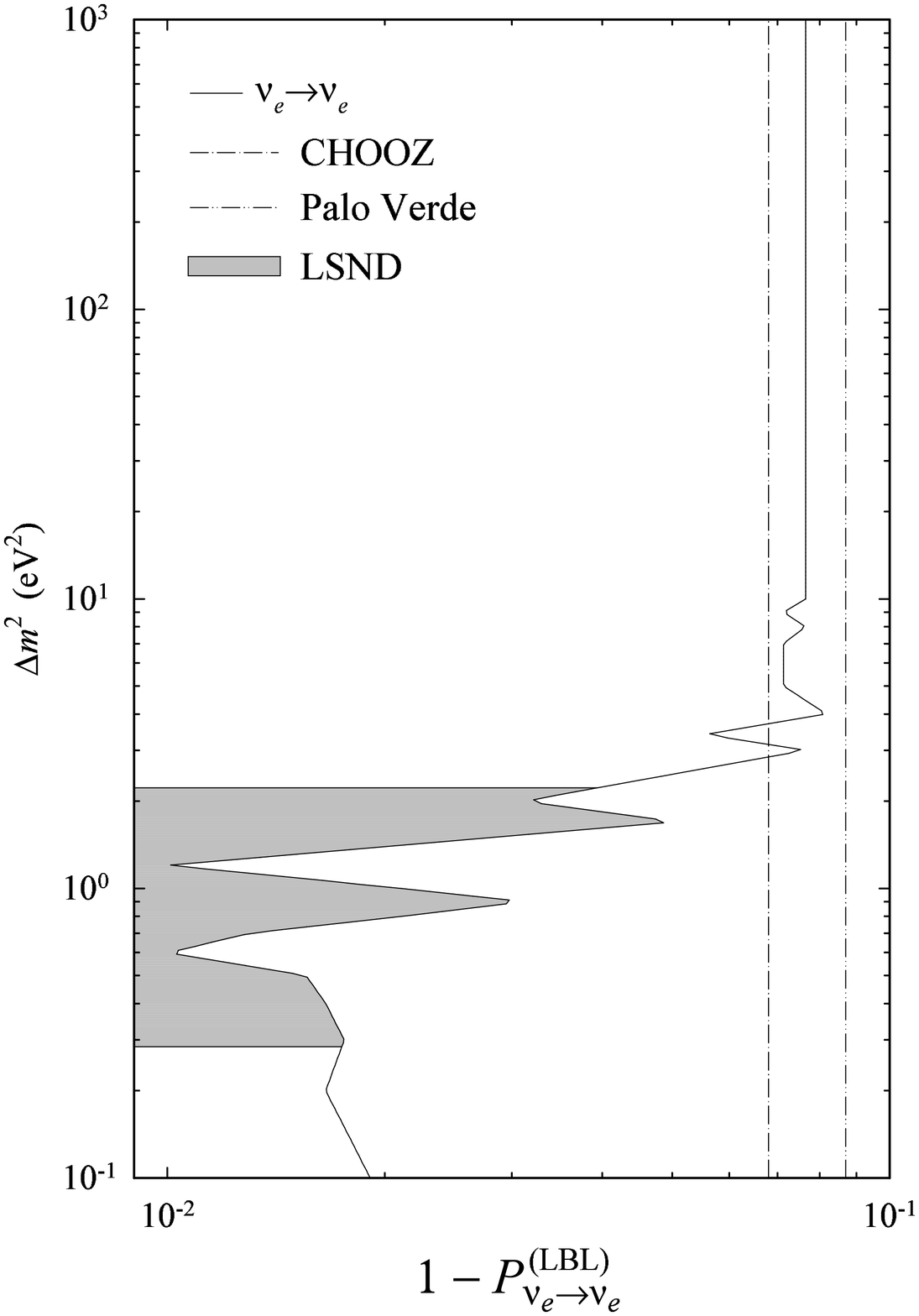,height=0.95\textheight}}
\end{center}
\end{minipage}
\begin{center}
\Large Figure~\ref{fig1}
\end{center}

\newpage

\begin{minipage}[p]{0.95\textwidth}
\begin{center}
\mbox{\epsfig{file=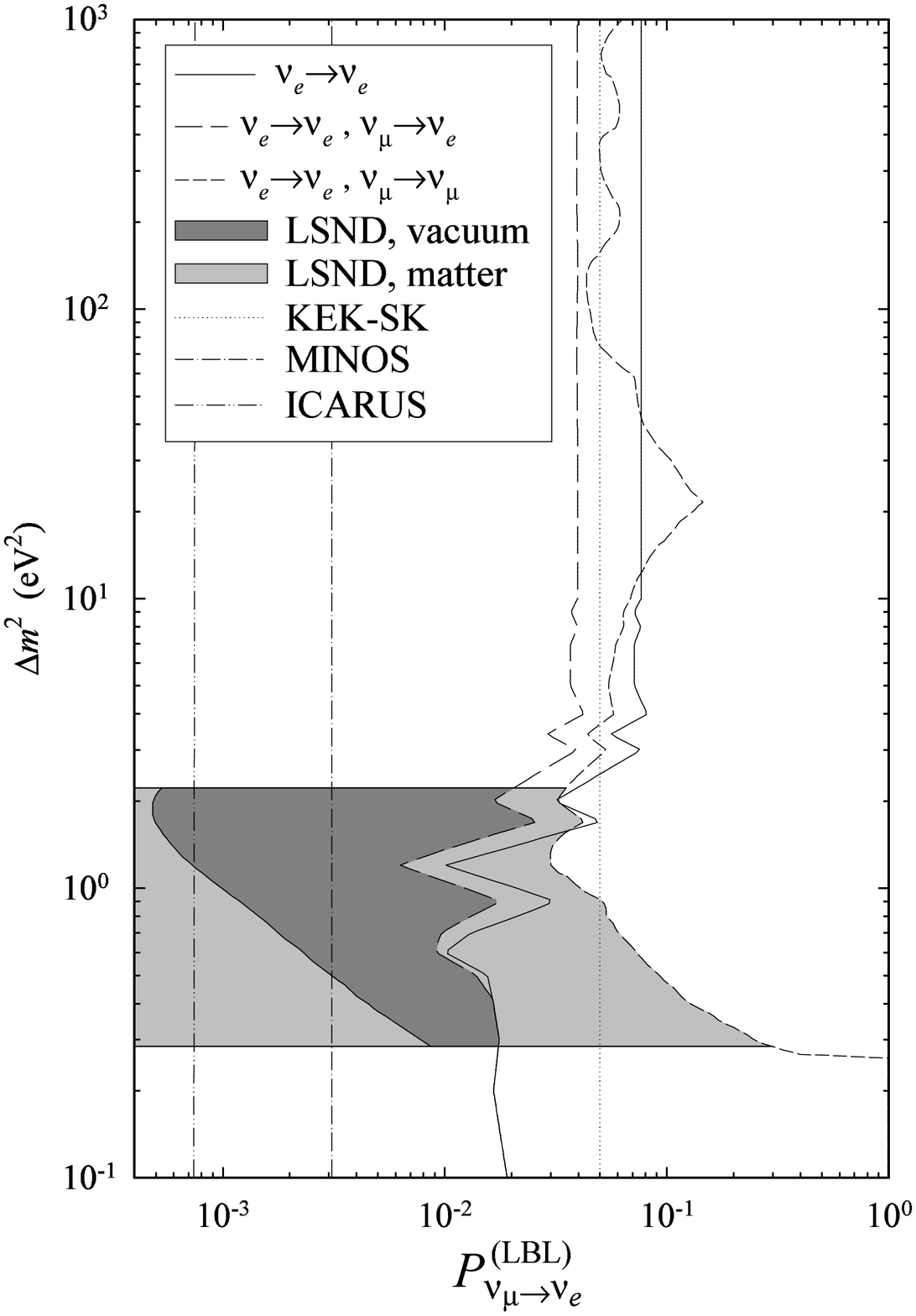,height=0.95\textheight}}
\end{center}
\end{minipage}
\begin{center}
\Large Figure~\ref{fig2}
\end{center}

\newpage

\begin{minipage}[p]{0.95\textwidth}
\begin{center}
\mbox{\epsfig{file=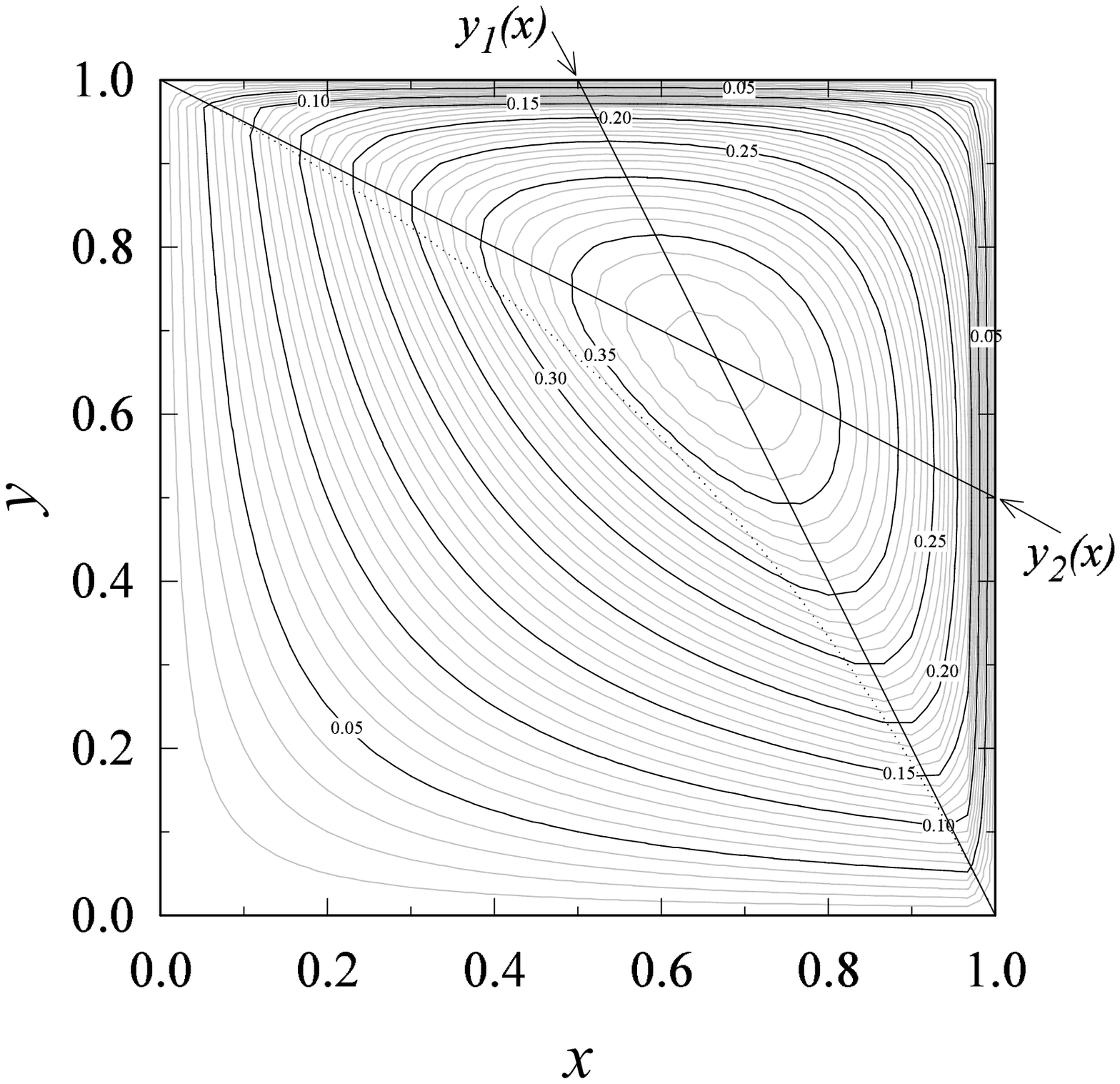,height=0.95\textheight}}
\end{center}
\end{minipage}
\begin{center}
\Large Figure~\ref{fig3}
\end{center}

\newpage

\begin{minipage}[p]{0.95\textwidth}
\begin{center}
\mbox{\epsfig{file=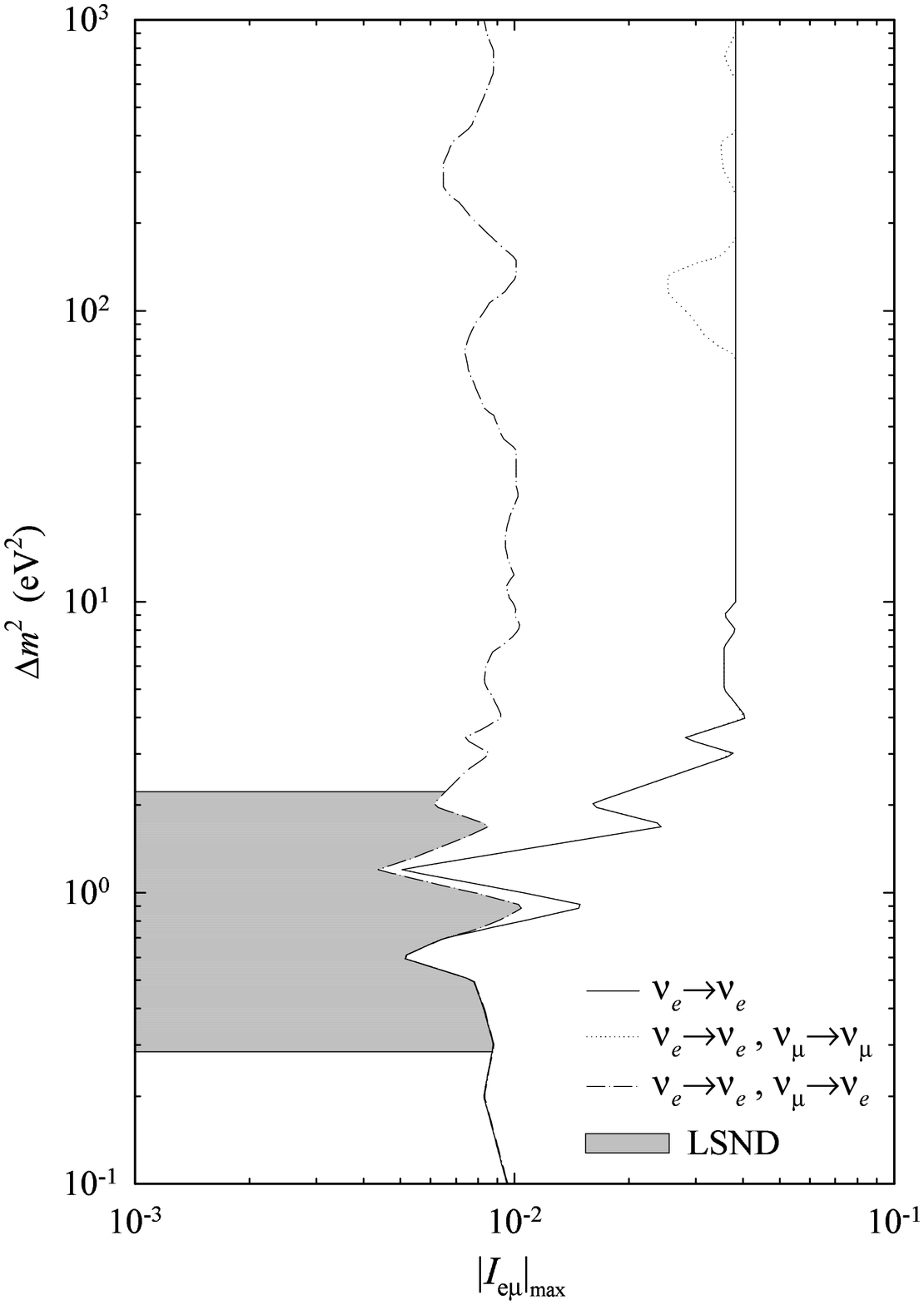,height=0.95\textheight}}
\end{center}
\end{minipage}
\begin{center}
\Large Figure~\ref{fig4}
\end{center}

\newpage

\begin{minipage}[p]{0.95\textwidth}
\begin{center}
\mbox{\epsfig{file=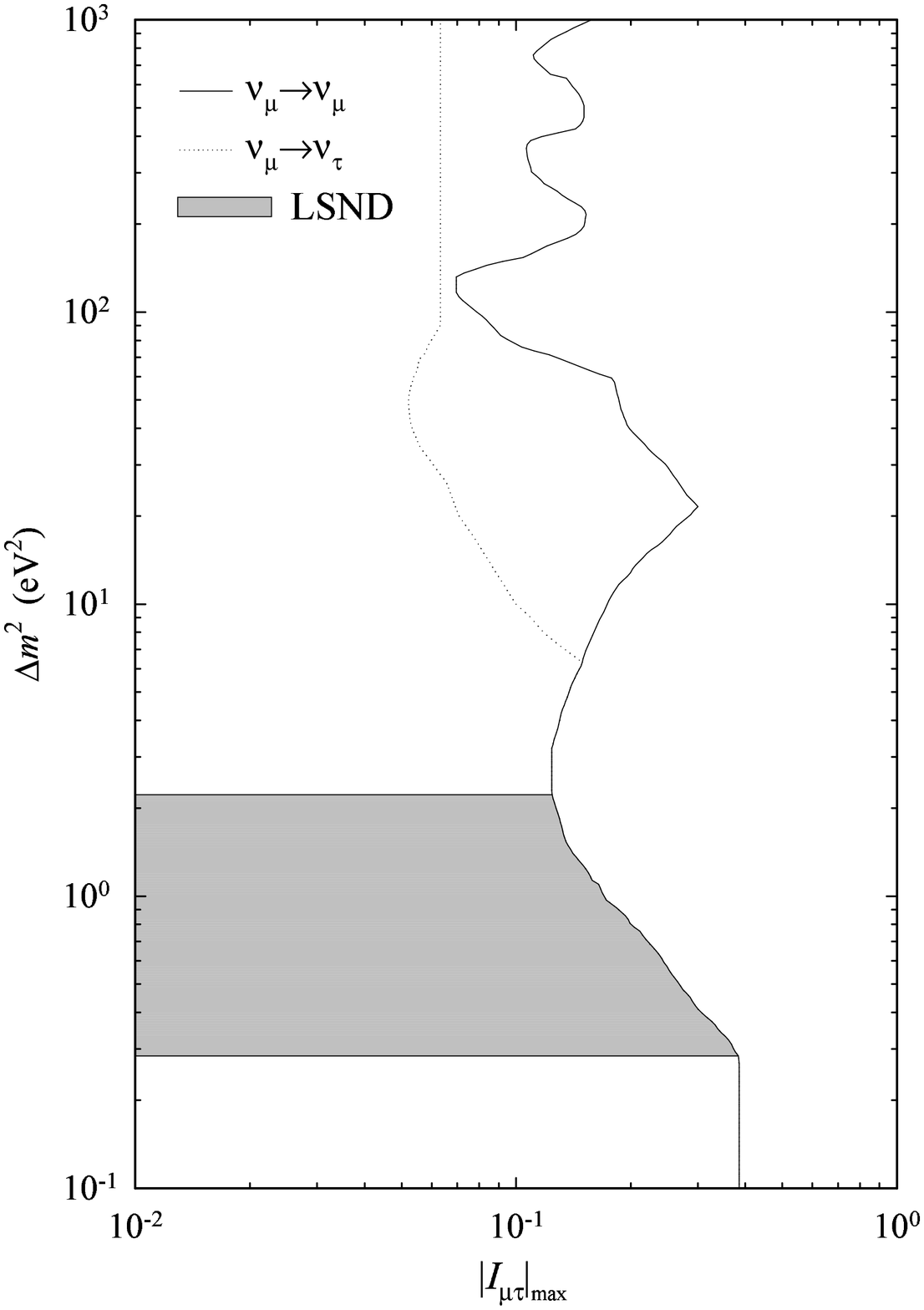,height=0.95\textheight}}
\end{center}
\end{minipage}
\begin{center}
\Large Figure~\ref{fig5}
\end{center}

\end{document}